\documentclass[12pt]{article}
\usepackage{hyperref} 
\usepackage{amsmath}
\usepackage{mathrsfs} 
\usepackage{enumerate}
\usepackage{cite}
\numberwithin{equation}{section}
\begin{document}
\def\be{\begin{equation}}
\def\bea{\begin{eqnarray}} 
\def\ee{\end{equation}}
\def\eea{\end{eqnarray}}
\def\d{\partial}
\def\eps{\varepsilon}
\def\la{\lambda}
\def\b{\bigskip}
\def\nn{\nonumber \\}
\def\p{\partial}
\def\t{\tilde}
\def\h{{1\over 2}}
\def\be{\begin{equation}}
\def\bea{\begin{eqnarray}}
\def\ee{\end{equation}}
\def\eea{\end{eqnarray}}
\def\b{\bigskip}
\def\u{\uparrow}
\def\AA{\mathscr{A}}
\def\DD{\mathscr{D}}
\def\FF{\mathscr{F}}
\def\LL{\mathscr{L}}
\def\CC{\mathscr{C}}
\def\MM{\mathscr{M}}
\newcommand{\comment}[2]{#2}

\setcounter{tocdepth}{2}

\makeatletter
\def\blfootnote{\xdef\@thefnmark{}\@footnotetext}  
\makeatother

\begin{center}
{\LARGE Gravitational Waves in the Myers--Perry Geometry}
\\
\vspace{18mm}
{\bf   Oleg Lunin}

\vspace{14mm}
{\it
Department of Physics,\\ University at Albany (SUNY),\\ Albany, NY 12222, USA\\
}
\vskip 10 mm

\blfootnote{email: olunin@albany.edu}

\end{center}

\begin{abstract}

We analyze equations describing gravitational waves in the Myers--Perry and Gibbons--Lu--Page--Pope geometries with arbitrary rotation parameters. Assuming that at least one rotation parameter vanishes, we demonstrate full separability of equations for several polarizations of gravitational waves and analyze the resulting ODEs. We also construct some examples of separable solutions describing gravitational excitations of black holes with the maximal number of rotation parameters.

\b

\end{abstract}

\newpage

\tableofcontents

\newpage

\section{Introduction}

Black holes are important laboratories for studying classical and quantum gravity. Over the last decade, an enormous progress in understanding these objects has been made through direct observations of electromagnetic \cite{BHemObs} and gravitational \cite{BHgravObs} waves produced by them. Reconstruction of images from such measurements requires a detailed understanding of classical fields in the vicinity of black holes, and the relevant calculations are usually done numerically \cite{NumGR1w1,NumGR1w2,NumGR1w3,NumGR1w4,NumGR2w1,NumGR2w2,NumGR2w3,
NumGR2w4,NumGR2w5}. Excitations of  idealized isolated black holes have also been studied analytically \cite{TeukW1,TeukW2}, and over the last five decades, these classic results have played an important role in improving our understanding of classical scattering by compact astrophysical objects and quantum effects in the vicinity of black holes. Specifically, Teukolsky's equations \cite{TeukW1,TeukW2} have been used to study the properties of the Hawking radiation \cite{Hawking} emitted as gravitons, photons, and light fermions in four dimensions. On the other hand, the microscopic states contributing to black hole entropy \cite{Beken} are still not fully understood, and the most significant progress has been made in higher dimensions
within the framework of string theory \cite{StrVafa,BMPVw1,BMPVw2,5dBHw1,5dBHw2,5dBHw3}. This raises a natural question of extending  
Teukolsky's analysis to higher dimensions, and the goal of our article is to address this question. 

\bigskip

Studies of black holes in higher dimensions have a long history which started with a discovery of the Schwarzschild--Tangherlini solution \cite{Tang} and its generalization to the rotating case found by Myers and Perry (MP) \cite{MyersPerry}. Charged versions of higher dimensional black holes have also been constructed \cite{HorStromW1,HorStromW2,HorStromW3,SenW1,SenW2,SenW3,SenW4,SenW5,SenW6,BMPVw1,BMPVw2}, and they led to remarkable progress in understanding quantum properties of black holes, such as counting of microscopic states contributing to the Bekenstein--Hawking entropy \cite{StrVafa,BMPVw1,BMPVw2}, and in exploration of strongly coupled field theories \cite{5dRotAdScftW1,5dRotAdScftW2,5dRotAdScftW3,5dRotAdScftW4,5dRotAdScftW5} in the framework of the AdS/CFT correspondence \cite{AdSCFTw1,AdSCFTw2,AdSCFTw3}. An extension of the MP geometry to systems with a negative cosmological constant was found by Gibbons--Lu--Page--Pope (GLPP) \cite{GLPP}, and various physical properties of this geometry have been analyzed in \cite{GLPPflwUpW1,GLPPflwUpW2,GLPPflwUpW3,GLPPflwUpW4}. Our article is dedicated to the study of gravitational waves in the MP and GLPP backgrounds. Light excitations of higher--dimensional black holes have already been extensively analyzed in the past. In the static case, scattering of various fields from charged black holes was studied in \cite{PreAdSCFTw1,PreAdSCFTw2,PreAdSCFTw3,PreAdSCFTw4,PreAdSCFTw5,PreAdSCFTw6,
PreAdSCFTw7}, and this work stimulated the discovery of AdS/CFT correspondence \cite{AdSCFTw1,AdSCFTw2,AdSCFTw3}. The success in computing scattering amplitudes and cross sections in static backgrounds \cite{PreAdSCFTw1,PreAdSCFTw2,PreAdSCFTw3,PreAdSCFTw4,PreAdSCFTw5,PreAdSCFTw6,
PreAdSCFTw7} relied on expanding all fields in spherical harmonics. Interestingly, for gravitons, even this problem turned out to be nontrivial, and a full study of gravitational excitations of static backgrounds was carried out in a very impressive series of articles \cite{KodamaOne,5DGravWaveW1,5DGravWaveW2,5DGravWaveW3,5DGravWaveW4,5DGravWaveW5,
5DGravWaveW6}. 

In the rotating case, the situations is more complicated. In contrast to the $d$--dimensional static black holes solutions, with have $SO(d-1)\times U(1)$ isometries responsible for full separability of equations for all dynamical excitations in spherical coordinates, MP and GLPP black holes have only  $[\frac{d+1}{2}]$ isometry directions, leading to the same number of conserved quantities. 
Nevertheless, equations of motion for various excitations of the MP and GLPP geometries turn out to be fully separable \cite{PreKubW1,PreKubW2,PreKubW3,Kub1w1,Kub1w2,Kub1w3,Kub1w4,Kub1w5,Kub1w6} due to additional conserved quantities associated with Killing and Killing--Yano tensors \cite{YanoW1,YanoW2,YanoW3}. The hidden symmetries responsible for this separation were first discovered for the Kerr black hole in \cite{Carter1w1,Carter1w2,Carter1w3}, and in the last three decades they have been extended to rotating black holes in an arbitrary number of dimensions in \cite{PreKubW1,PreKubW2,PreKubW3,Kub1w1,Kub1w2,Kub1w3,Kub1w4,Kub1w5,Kub1w6,Kub2w1,
Kub2w2,Kub2w3,Kub3w1,
Kub3w2,Kub3w3,KubRev}, including some geometries carrying charges \cite{StrYanoW1,StrYanoW2,StrYanoW3,ChLKill}\footnote{As a complement to this general construction relying on the Killing--Yano tensors, separation of variables for various fields in charged geometries in $d>4$ has also been explored on a case--by--case basis in \cite{LarsenW1,LarsenW2,LarsenW3,LarsenW4,LTianW1,LTianW2,LTianW3,DeLu24}.}. 
In four dimensions, Killing tensors were first used to separate variables in  equations for probe particles and scalar fields in \cite{Carter1w1,Carter1w2,Carter1w3}, and subsequent Teukolsky's construction for fields with higher spins \cite{TeukW1,TeukW2} also implicitly relied on such objects. Specifically, Teukolsky demonstrated separation of variables in projections of gauge--invariant quantities, such as the field strength for Maxwell field and the Weyl tensor for gravity, on the eigenvectors of Killing tensors also known as the Newman--Penrose tetrads \cite{NewmPenr}. In higher dimensions, separability of equations for probe particles and scalars 
 in the MP and GLPP backgrounds 
was shown in \cite{PreKubW1,PreKubW2,PreKubW3,Kub1w1,Kub1w2,Kub1w3,Kub1w4,Kub1w5,Kub1w6}, and it was later extended to the Maxwell field \cite{LMaxw}, massive vectors \cite{Proca}, and higher forms \cite{LBfield}.  Again, projections on eigenvectors of Killing--Yano tensors played a crucial role in these constructions, but such projections involved gauge potentials rather than fields strengths. Motivated by these successes, in this article we will study projections of metric components rather than curvatures on the preferred frames (the higher--dimensions analogs of the Newman--Penrose tetrads) and demonstrate their separability.

Although full integrability of equations for gravitational waves on MP and GLPP backgrounds has not been demonstrated yet, several studies have been performed in special situations on a case--by--case basis. Specifically, several modes for black holes with rotation in only one plane were analyzed in \cite{KodamaRotOneAw1,KodamaRotOneAw2}, and similar studies for equal rotations in all planes were carried out in \cite{GWequalA}. Separability of equations for gravitons in the near--horizon limit of the MP geometry, where the symmetry is enhanced, was investigated in \cite{GWnearHor}, and separability of gravitational waves in five dimensions was analyzed in \cite{AllDimOtherW1,AllDimOtherW2,AllDimOtherW3}. 
More generally, as the first step towards extending Teukolsky's construction to higher dimensions, article \cite{NPallD} made a significant progress in generalizing the Newman--Penrose formalism beyond four dimensions, but so far this procedure has worked only for some special modes. Our approach is complementary to these developments. Motivated by successes in separating variables in equations for vector fields and higher forms \cite{LMaxw,Proca,LBfield}, we impose an ansatz not on gauge invariant variables used in \cite{KodamaOne,5DGravWaveW1,5DGravWaveW2,5DGravWaveW3,5DGravWaveW4,5DGravWaveW5,
5DGravWaveW6,NPallD}, but on certain projections of the metric components. Also, we set one rotation parameter to zero while keeping keep all other ones arbitrary, thus extending the construction of \cite{KodamaRotOneAw1,KodamaRotOneAw2} and complementing the one of \cite{GWequalA}. In section \ref{SecHair} we discuss some approaches to relaxing the requirement of a vanishing rotation parameter. 

\bigskip

This paper has the following organization. In section \ref{SecModes} we briefly review classification of gravitational waves for systems reduced on spheres. Such reduction naturally arises for the MP and GLPP geometries once some rotation parameters are set to zero. If the background geometry can be split into a sphere with some warp factor and the remaining directions (the base), then gravitational waves naturally split into scalar, vector and tensor modes. In the remaining part of the article we focus on excitations described by vectors and scalars on the base. In section \ref{SecVector} we study vector perturbations, and section \ref{SecScalar} is dedicated to the analysis of the scalar ones. In section \ref{SecHair} we also describe a promising approach to gravitational waves on geometries with a maximal number of non--zero rotation parameters, but unfortunately only some special solutions for this case are constructed. Given the technical nature of some results obtained in sections \ref{SecVector}-\ref{SecScalar}, we present a short summary of our findings in section \ref{SecSumry}. Some supplementary material is presented in the appendices.

\section{Classification of gravitational waves}
\label{SecModes}

In this article we focus on gravitational waves in black hole backgrounds with fewer than the maximal number of rotations. As reviewed in the Appendix \ref{AppMP}, a black hole in $d$ dimensions can have up to ${\bar n}=\left[\frac{d-1}{2}\right]$ rotation parameters\footnote{We denote the maximal number of rotation parameters by ${\bar n}$, and we reserve $n$ for the number of non--cyclic angular coordinates in the MP and GLPP geometries. One has $n={\bar n}$ in even dimensions and ${\bar n}=n+1$ otherwise.}. If $p$ such parameters are set to zero, then the metric takes the form\footnote{The relation between $s$ and $p$ is different in even ($s=2p$) and odd ($s=2p-1$) number of dimensions $d$, and it will be discussed in detail in section \ref{SecSubProca}.}
\bea\label{SpaceSplit}
ds^2={\bar g}_{\mu\nu}dx^\mu dx^\nu +[f(x)]^2 d\Omega_{s}^2,\quad 
d\Omega_{s}^2=q_{ab}dy^a dy^b\,,
\eea
where the first part describes the $(d-s)$--dimensional base space, and the second part describes an 
$s$--dimensional sphere. In this article we will denote the metric on the base by ${\bar g}_{\mu\nu}$ 
and the metric in the full $d$--dimensional space by $g_{MN}$. Classification of gravitational waves on spaces (\ref{SpaceSplit}) was developed in 
\cite{KodamaOne,5DGravWaveW1,5DGravWaveW2,5DGravWaveW3,5DGravWaveW4,5DGravWaveW5,
5DGravWaveW6}\footnote{See also \cite{KodamaExtraReviewW1,KodamaExtraReviewW2} for nice reviews of this formalism.}, and excitations naturally split into three decouples classes:
\begin{enumerate}[(a)]
\item {\bf Scalar modes on the sphere.}\\
The excitations are built from the scalar spherical harmonics $S(y)$, and the perturbations are given by
\bea\label{SetUpScal}
\hskip -0.3cm
ds_{\eps}^2=\eps\left[S h_{\mu\nu}dx^\mu dx^\nu+2(\d_a S)h^{(S)}_\mu dx^\mu dy^a+
h^{(1)}_SS d\Omega_{2s-1}^2+h^{(2)}_S({\hat\nabla}_a {\hat\nabla}_b S)dy^a dy^b\right],
\eea
where $(h_{\mu\nu},h_\mu,h^{(1)},h^{(2)})$ are functions of the coordinates parameterizing the base, and $S$ are scalar spherical harmonics defined by
\bea
({\hat\nabla}^2+\la_S) S=0\,.
\eea
Some properties of scalar, vector, and tensor spherical harmonics are discussed in the Appendix \ref{AppSphHarm}.

\item {\bf Vector modes on the sphere.} \\
The excitations are built from the vector spherical harmonics $V_a(y)$, and the perturbations are given by
\bea\label{ClasVec1}
ds_{\eps}^2=\eps\left[2 V_a h^{(V)}_\mu dx^\mu dy^a+
h_V({\hat\nabla}_a V_b+{\hat\nabla}_b V_a)dy^a dy^b\right],
\eea
where $(h_\mu,h_V)$ are functions of the coordinates on the base, and $V_a$ are vector spherical harmonics defined by
\bea\label{ClasVec2}
({\hat \nabla}^2+\la_V) V_a=0,\quad {\hat \nabla}_a V^a=0.
\eea
\item {\bf Tensor modes on the sphere.} \\
The excitations are built from the tensor spherical harmonics $T_{ab}(y)$, and the perturbations are given by
\bea\label{TensoModeDef}
ds_{\eps}^2=\eps h_T T_{ab}dy^a dy^b\,,
\eea
where $h_T$ is a function of the coordinates on the base, and $T_{ab}$ are tensor spherical harmonics defined by
\bea\label{TensoModeDef2}
({\hat \nabla}^2+\la_T) T_{ab}=0,\quad {\hat \nabla}_a T^{ab}=0,\quad q_{ab}T^{ab}=0.
\eea
Tensor modes are present only of the dimension of the sphere is greater than two. 
\end{enumerate}
In this article we will focus only on vector and tensor modes and demonstrate separability of the equations for gravitational waves in the Myers--Perry (MP) and  Gibbons--Lu--Page--Pope (GLPP) geometries \cite{MyersPerry,GLPP} with fewer than $\bar n$ rotational parameters. Specifically, in section \ref{SecVector}  we will use the insights from separability of Maxwell's equations in the MP and GLPP backgrounds \cite{MyersPerry} to solve the equations for vector excitations, and in section \ref{SecScalar} we will analyze the scalar equation, which is generically produced by the tensor modes in partially rotating black holes \cite{KodamaOne,5DGravWaveW1,5DGravWaveW2,5DGravWaveW3,5DGravWaveW4,5DGravWaveW5,
5DGravWaveW6}.
The scalar modes contain a tensor on the base space, so equations for them are as complicated as the ones for the generic gravitational waves on the MP backgrounds with ${\bar n}$ rotations, and analysis of such perturbations is beyond the scope of this article.

\section{Vector modes of partially rotating black holes}
\label{SecVector} 

The main motivation for this article comes from the desire to construct gravitational waves in 
the most general neutral rotating black holes with ${\bar n}=\left[\frac{d-1}{2}\right]$ rotation parameters. The solution to this problem seems to be beyond reach, although some progress in this direction is discussed in the next section, so here we will focus on excitations of black holes with fewer than ${\bar n}$ rotations. The most general geometry of this kind has $({\bar n}-1)$ rotational parameters, and sections \ref{SecVecOdd} and \ref{SecVecEven} will be dedicated to the study of vector modes in this case. Vector modes in backgrounds with fewer than $({\bar n}-1)$ rotations will be discussed in sections \ref{SecSubProca} and \ref{SecSubStatic}. 

\bigskip

If one of the rotation parameters is set to zero, then the Myers--Perry geometry \cite{MyersPerry} has the form
\bea\label{MetrFact}
ds^2={\bar g}_{\mu\nu}dx^\mu dx^\nu+f^2 d\psi^2,
\eea
and the cyclic coordinate $\psi$ does not appear in the first part of the metric. In this case, the vector excitations (\ref{ClasVec1})--(\ref{ClasVec2}) reduce to 
\bea\label{PertVec1}
ds_\eps^2=2 h_{\mu\psi}dx^\mu d\psi,
\eea
where $h_{\mu\psi}$ does not depend on the $\psi$ coordinate. Rather than using the classification of modes on $S^s$ \cite{KodamaOne,5DGravWaveW1,5DGravWaveW2,5DGravWaveW3,5DGravWaveW4,5DGravWaveW5,
5DGravWaveW6} reviewed in the previous section and applying it to $s=1$, one can just focus on the most general perturbations of the geometry (\ref{MetrFact}) that do not depend on the coordinate
$\psi$. Since the geometry has a symmetry under reflection of $\psi$, such excitations split into two decoupled classes, $h_{\psi\mu}$ and the rest, and we will focus on the first option given by (\ref{PertVec1}). 
In other words, we will consider a perturbed geometry
\bea\label{VectPertDef}
ds^2={\bar g}_{\mu\nu}dx^\mu dx^\nu+f^2 (d\psi+\eps A_\mu dx^\mu)^2,
\eea
and look at the Einstein's equations in the first order in $\eps$. Recalling the general procedure for dimensional reduction \cite{SenW1,SenW2,SenW3,SenW4,SenW5,SenW6}, applying it to the current case, and expanding the action to quadratic order in $\eps$, we find
\bea
S=\frac{1}{32\pi}\int d^{d-1} x d\psi\sqrt{-{\bar g}}f\left[R_{d-1}-
\frac{\eps^2}{4} f^2 F_{\mu\nu}F^{\mu\nu}-\frac{1}{4}\d_\mu f\d^\mu f^{-1}+
\frac{1}{4}\d_\mu f\d^\mu f^{-1}\right]\nonumber
\eea
The vector part of the action reads
\bea\label{VectPertDefAct}
S_{vec}=-\frac{\eps^2}{128\pi}\int d^{d-1} x d\psi\sqrt{-{\bar g}} f^3 F_{\mu\nu}F^{\mu\nu}
=-\frac{\eps^2}{128\pi}\int d^{d} x\sqrt{-{g}}f^2 F_{\mu\nu}F^{\mu\nu}
\eea
Therefore, gauge field $A_\mu$ satisfies the modified Maxwell's equations in the geometry (\ref{MetrFact})
\bea\label{MaxwDefrm}
\frac{1}{\sqrt{-g}}\d_\mu\left[\sqrt{-g} f^2 F^{\mu\nu}\right]=0.
\eea
Later in this section we will construct separable solutions of the system (\ref{MaxwDefrm}), but before doing that, let us comment on extending these equations to geometries with a negative cosmological constant. Such generalizations of the Myers--Perry black holes were obtained in \cite{GLPP}, and these GLPP solutions are reviewed in the Appendix \ref{SecSubMP}. To study perturbations of the GLPP geometries, we start with the metric (\ref{VectPertDef}) and impose the Einstein's equations
\bea\label{EinstLamb}
R_{MN}=\Sigma g_{MN}\,,
\eea
where $\Sigma$ is proportional to the cosmological constant $\Lambda$. Expanding both sides of equation (\ref{EinstLamb}) to the first order in $\eps$, and assuming that the unperturbed metric satisfies this equation, we find
\bea
R_{MN}=R^{(0)}_{MN}+\eps R^{(1)}_{MN},&& R^{(1)}_{\psi\mu}=
-\frac{1}{2\sqrt{-g}}g_{\mu\nu}\d_\sigma\left[\sqrt{-g} f^2 F^{\sigma\nu}\right]+
\Sigma f^2 A_\mu\nn
&& R^{(1)}_{\mu\nu}=0,\quad R^{(1)}_{\psi\psi}=0\\
g_{MN}=g^{(0)}_{MN}+\eps g^{(1)}_{MN},&& g^{(1)}_{\psi\mu}=f^2 A_\mu,\quad
g^{(1)}_{\mu\nu}=0,\quad g^{(1)}_{\psi\psi}=0.\nonumber
\eea
Therefore, equation (\ref{EinstLamb}) for linear perturbations reduces to the generalized Maxwell's equation (\ref{MaxwDefrm}) even in the presence of the cosmological constant. 

Let us now demonstrate separability of the generalized Maxwell's equation (\ref{MaxwDefrm}) in the MP and GLPP backgrounds. Given the structure of these geometries, it is convenient to have separate discussions of even and odd--dimensional cases.  

\subsection{Reduction on a circle in odd dimensions}
\label{SecVecOdd}

The MP and GLPP geometries in $d=2n+3$ dimensions are reviewed in Appendix \ref{SecSubMP}. Here we just observe that the metrics can be written in terms of convenient separable frames (\ref{GenFramesOddD}). If one of the rotation parameters is set to zero, then the frames become
\bea\label{GenOddTnc}
&&\hskip -1cm
\left[m_\pm^{(0)}\right]^\mu\d_\mu=\frac{R}{\sqrt{\Delta}}\left\{\frac{Q_r\Delta}{R}\d_r\pm \frac{1}{Q_r}\left[\d_t-
\sum_k^n\frac{a_k}{r^2+a_k^2}\d_{\phi_k}\right]\right\},\ \Delta=R-Mr^2,\\
&&\hskip -1cm\left[m_\pm^{(j)}\right]^\mu
\d_\mu=\sqrt{{H_j}}\left\{Q_j\d_{x_j}\pm \frac{i}{Q_j}\left[\d_t-\sum_k^n\frac{a_k}{a_k^2-x_j^2}\d_{\phi_k}
\right]\right\}\,,\quad
n^\mu\d_\mu=-\d_{\psi}\,.\nonumber
\eea
Here functions $(R,H_i)$ are defined by
\bea\label{RHreduced}
R=r^2\prod_k^{n}(r^2+a^2_k),\quad H_i=-x_i^2\prod_k^{n}(a_k^2-x_i^2),
\eea
and factors $(Q_r,Q_j)$ defined by (\ref{QfactDef}) contain information about the cosmological parameter $L$. To shorten some formulas below, we will occasionally use $l_\pm^\mu$ instead of 
$\left[m_\pm^{(0)}\right]^\mu$ and denote the differential operator corresponding to vector $V^\mu$ by ${\hat V}$. For example,
\bea
{\hat l}_\pm\equiv {\hat m}_\pm^{(0)}\equiv \left[m_\pm^{(0)}\right]^\mu\d_\mu,\quad
{\hat m}_\pm^{(j)}\equiv\left[m_\pm^{(j)}\right]^\mu\d_\mu,\quad {\hat n}\equiv n^\mu\d_\mu\,.
\eea
The MP geometry has $L=0$ and $Q_r=Q_j=1$.  

Separation of variables in equations for various excitations of the MP and GLPP geometries relies on the crucial feature of the reduced frames (\ref{GenOddTnc}): the components  $l_\pm^\mu$ depend only on the radial coordinate, the components $\left[m_\pm^{(j)}\right]^\mu$  depend only on $x_j$, and the components of $n^\mu$ are constants. The inverse metric has the form
\bea\label{GenOddMetrTnc}
g^{\mu\nu}\d_\mu\d_\nu&=&\frac{1}{FR}\left[m_+^{(0)}\right]^\mu 
\left[m_-^{(0)}\right]^\nu\d_\mu\d_\nu+
\sum_{j=1}^n \frac{1}{x_j^2 d_j}{\left[m_+^{(j)}\right]^\mu\left[m_-^{(j)}\right]^\mu\d_\mu\d_\nu}\\
&&+
\left[\frac{\prod a_i}{r\prod x_k}\right]^2n^\mu n^\nu\d_\mu\d_\nu\,,\nonumber
\eea
where $d_j$ and $FR$ are given by (\ref{MiscElliptic}) and (\ref{DelktaOdd}). 
In contrast to (\ref{GinvOddApp}), the product of rotation parameters in the second line of (\ref{GenOddMetrTnc}) does not contain $a_n$, which is currently set to zero.
As expected, the metric (\ref{GenOddMetrTnc}) has the form (\ref{MetrFact}) with 
\bea\label{PrefacOdd}
f=\frac{r\prod x_k}{\prod a_i}\,.
\eea
Note that the product in the denominator contains only $(n-1)$ non--zero rotational parameters, and function $f$ is multiplicatively separable in $(r,x_1,\dots x_{n-1})$. 

Motivated by (\ref{MaxwDefrm}), we consider a generalized Maxwell's equation for an arbitrary separable function 
$S$:
\bea\label{MaxwDefrmS}
\frac{1}{\sqrt{-g}}\d_\mu\left[\sqrt{-g}S F^{\mu\nu}\right]=0.
\eea
In the standard case, where $S=1$, separability of this equation in the MP and GLPP backgrounds was demonstrated in \cite{LMaxw}. Specifically, it was shown that imposing an ansatz
\bea\label{AnstzMaxwGW}
l_{\pm}^\mu A_{\mu}=\left[\frac{r}{r\mp i\mu}+\zeta\right]{\hat l}_\pm \Psi,\quad
{m}^{(j)\mu}_\pm  A_{\mu}=\left[\frac{x_j}{x_j\pm \mu}+\zeta\right]{\hat m}^{(j)}_\pm{\Psi}\,\quad
n^\mu A_\mu=\zeta {\hat n}{\Psi}
\eea
with a separable function $\Psi$,
\bea\label{PsiSeparDef}
\Psi=E\Phi(r)\left[\prod X_i(x_i)\right],\qquad E=e^{i\omega t+i\sum n_i\phi_i}\,,
\eea
and substituting the resulting gauge field into Maxwell's equations, one arrives at a system of ODEs which contains $(n-1)$ separation constants in addition to $\mu$. Parameter $\zeta$ describes a pure gauge. It was later observed in \cite{KubMaxwW1,KubMaxwW2} that one recovers the Lorenz gauge by setting $\zeta=-1$. 

Let us now impose the ansatz (\ref{AnstzMaxwGW}) with a separable function $\Psi$ on the gauge field and substitute the result into the modified Maxwell's equations (\ref{MaxwDefrmS}). Calculations show that equations are consistent if and only if function $S$ has the form
\bea\label{SfactOdd}
S=(r\prod x_k)^\nu
\eea
with an arbitrary parameter $\nu$. The resulting ODEs read
\bea\label{ODEoddMgen}
&&\frac{D_r}{Sr}\frac{\d}{\d r}\left[\frac{SQ_r^2\Delta}{rD_r}\frac{\d\Psi}{\d r}\right]+\left\{\frac{2\Lambda}{D_r}+\nu\Lambda+\frac{R^2 W_r^2}{r^2Q_r^2\Delta}+P_{n-2}[r^2]D_r\right\}\Psi=0,\\
&&\frac{D_j}{Sx_j}\frac{\d}{\d x_j}\left[\frac{SQ_j^2H_j}{x_jD_j}\frac{\d\Psi}{\d x_j}\right]+\left\{\frac{2\Lambda}{D_j}+\nu\Lambda-\frac{H_j W_j^2}{x_j^2Q_j^2}+P_{n-2}[-x_j^2]D_j\right\}\Psi=0,\nonumber
\eea
where $P_{n-2}$ is an arbitrary polynomial of degree $(n-2)$, and functions $(D_r,D_j,W_r,W_j)$ and constants $(\Omega,\Lambda)$ are given by
\bea\label{AAdef}
&&W_j=\omega-\sum_k\frac{n_k a_k}{a_k^2-x_j^2},\quad 
W_r=\omega-\sum_k\frac{n_k a_k}{a_k^2+r^2},\quad
D_j=1-\frac{x_j^2}{\mu^2},\quad D_r=1+\frac{r^2}{\mu^2}\nn
&&\Omega=\omega-\sum_k\frac{n_ka_k}{\Lambda_k},\quad 
\Lambda=\frac{\Omega}{\mu^3}
\prod \Lambda_k,\quad \Lambda_i=(a_i^2-{\mu^2}).
\eea
As expected, the gauge parameter $\zeta$ did not appear in the system (\ref{ODEoddMgen}). The solutions are specified by $2n+1$ free parameters: 
\begin{itemize}
\item $n+1$ parameters $(\omega,n_i)$;
\item $(n-1)$ free coefficients of the polynomial $P_{n-2}$;
\item constant $\mu$.
\end{itemize}
This gives $2n+1=d-2$ constants which guarantee the most general separation on a $(d-1)$--dimensional base that does not include the $\psi=\phi_n$ direction. 

Recalling that the metric (\ref{GenOddMetrTnc}) has the form (\ref{MetrFact}) with a prefactor (\ref{PrefacOdd}), we conclude that the vector components of gravitational waves satisfy the modified Maxwell's equations (\ref{MaxwDefrmS}), and factor $S$ is given by (\ref{SfactOdd}) with $\nu=2$. Therefore, gravitational perturbations are separable, and function $\Psi$ satisfied the system of ODEs (\ref{ODEoddMgen}). For future reference it is instructive to summarize the nontrivial projections of the waves on the frames:
\bea\label{AnstzMaxw}
l_{\pm}^\mu n^\nu h_{\mu\nu}=S\left[\frac{r}{r\mp \mu}+\zeta\right]{\hat l}_\pm \Psi,\quad
m_{\pm}^{(j)\mu}  n^\nu h_{\mu\nu}=
S\left[\frac{x_j}{x_j\pm \mu}+\zeta\right]{\hat m}^{(j)}_\pm{\Psi}\,.
\eea
All projections not listed here vanish\footnote{In particular, the last projection in (\ref{AnstzMaxwGW}) does not contribute since the generalized Maxwell's equation (\ref{MaxwDefrm}) was obtained under two  assumption: $\psi$--independence of function $\Psi$ and $A_\psi=0$. Then the last projection in (\ref{AnstzMaxwGW}) gives an identity $0=0$.}. The equations for function $\Psi$ are given by 
(\ref{ODEoddMgen}) with
\bea\label{OddNuParam}
S=\left[r\prod x_k\right]^2,\quad \nu=2.
\eea
This concludes our discussion of vector modes in odd dimensions. 

\subsection{Reduction on a circle in even dimensions}
\label{SecVecEven}

The MP and GLPP geometries in $d=2n+2$ dimensions are reviewed in the Appendix \ref{SecSubMP}. When all $n$ rotational parameters are turned on, the reduced special frames are given by (\ref{Mframes}), and in contrast to the odd dimensional case (\ref{GenOddTnc}), this time there is no counterpart of the vector $n^\mu$. Without loss of generality, the ellipsoidal coordinates $x_j$ can be assumed to take values in the ranges
\bea\label{EvenElptRange}
0\le x_n\le a_n\le \dots \dots \le x_1\le a_1.
\eea
Let us now discuss the limiting cases of the MP and GLPP geometries for $a_n=0$. 

\bigskip
Given the ranges (\ref{EvenElptRange}), we conclude that the $a_n\rightarrow 0$ limit must be accompanied 
by sending $x_n$ to zero as well. To implement this, we write  $x_n=a_n y_n$ where $y_n$ takes values between zero and one. Then taking the $a_n\rightarrow 0$ limit in the frames (\ref{Mframes}), we find 
\bea
&&\hskip -1cm\left[m_\pm^{(0)}\right]^\mu\d_\mu=\frac{R}{\sqrt{\Delta}}\left\{\frac{Q_r\Delta}{R}\d_r\pm \frac{1}{Q_r}\left[\d_t-
\sum_{k<n}\frac{a_k}{r^2+a_k^2}\d_{\phi_k}\right]\right\},\quad 
\Delta=R-Mr,
\nn
&&\hskip -1cm\left[m_\pm^{(j)}\right]^\mu
\d_\mu=\sqrt{{H_j}}\left\{Q_j\d_{x_j}\pm \frac{i}{Q_j}\left[\d_t-\sum_{k<n}\frac{a_k}{a_k^2-x_j^2}\d_{\phi_k}
\right]\right\}\,,\\
&&\hskip -1cm\left[m_\pm^{(n)}\right]^\mu
\d_\mu=\sqrt{(1-y_n^2)\prod_{k<n}a_k^2}\left\{\d_{y_n}\pm i\left[-\frac{1}{1-y_n^2}\d_{\phi_n}
\right]\right\},\nonumber
\eea
where functions $(R,H_i)$ are still given by (\ref{RHreduced}). As already mentioned, there is no counterpart of the vector $n^\mu$. The inverse metric reads
\bea\label{GenEvenMetrTnc}
g^{\mu\nu}\d_\mu\d_\nu&=&\frac{1}{FR}\left[m_+^{(0)}\right]^\mu 
\left[m_-^{(0)}\right]^\nu\d_\mu\d_\nu+
\sum_{j=1}^{n-1} \frac{1}{d_j}{\left[m_+^{(j)}\right]^\mu\left[m_-^{(j)}\right]^\mu\d_\mu\d_\nu}
\\
&&+\frac{1}{r^2\prod x_j^2}{\left[m_+^{(n)}\right]^\mu\left[m_-^{(n)}\right]^\mu\d_\mu\d_\nu}\,.
\nonumber
\eea
It is easy to see that the last line corresponds to a decoupled metric of the two--sphere with a warp factor\footnote{Recall that the general definition of such a warp factor is given by  (\ref{SpaceSplit}).}
\bea
f_{S^2}=\frac{r\prod x_k}{\prod a_i}\,.
\eea
This description will become important in section \ref{SecSubProca}. Alternatively, one may perform a reduction of the metric (\ref{GenEvenMetrTnc}) on $\phi_n$ at the expense of making some $SO(3)$ isometries implicit. This allows a uniform treatment of even and odd dimensions since the metric (\ref{GenEvenMetrTnc}) can be written in the form (\ref{VectPertDef}) with 
\bea\label{FormFactEven}
\psi=\phi_n,\quad f=\frac{r\prod x_k}{\prod a_i}\sqrt{1-y_n^2}\,.
\eea
The last expression can be made more symmetric and more similar to (\ref{PrefacOdd}) if one introduces a new coordinate 
$z_n=\sqrt{1-y_n^2}$:
\bea\label{PrefacEven}
f=\frac{r\prod x_k}{\prod a_i}z_n,\quad 
\left[m_\pm^{(n)}\right]^\mu
\d_\mu=-\sqrt{\prod_{k<n}a_k^2}\left\{\sqrt{1-z_n^2}\d_{z_n}\pm \frac{i}{z_n}\d_{\phi_n}\right\}
\eea
After presenting this brief summary of the MP and GLPP metrics, we now proceed to solving the modified Maxwell's equation (\ref{MaxwDefrm}) which came from reduction of gravitational waves on a circle. 

\bigskip

We begin with looking at equation (\ref{MaxwDefrm}) in the non--degenerate MP or GLPP geometries with frames (\ref{Mframes}) and imposing the ansatz
\bea\label{AnstzMaxwEvenOne}
\hskip -0.5cm
l_{\pm}^\mu A_{\mu}=\left[\frac{r}{r\mp i\mu}+\zeta\right]{\hat l}_\pm \Psi,\ \ 
m_{\pm}^{(j)\mu}  A_{\mu}=\left[\frac{x_j}{x_j\pm \mu}+\zeta\right]{\hat m}^{(j)}_\pm{\Psi}\,,
\eea
which has been successfully used to separate the Maxwell and the Proca equations in the past \cite{LMaxw,Proca}. Here $\Psi$ is a separable function that has the form (\ref{PsiSeparDef}). Substituting the resulting gauge field into the modified Maxwell's equations (\ref{MaxwDefrm}), one finds that function $S$ must have the form
\bea\label{SfactEvenNG}
S=(r\prod x_k)^\nu\,,
\eea
and function $\Psi$ satisfies a system of ODEs
\bea\label{EvenDimGenSpin}
&&\frac{D_r}{S}\frac{\d}{\d r}\left[\frac{S Q_r^2\Delta}{D_r}\frac{\d\Psi}{\d r}\right]-\left\{\frac{2\Lambda}{D_r}-
\frac{R^2 W_r^2}{Q_r^2\Delta}+(\nu-1)\Lambda +P_{n-2}[r^2]  D_r\right\}\Psi=0,
\\
&&\frac{D_j}{S}\frac{\d}{\d x_j}\left[\frac{S Q_j^2 H_j}{D_j}\frac{\d\Psi}{\d x_j}\right]+\left\{\frac{2\Lambda}{D_j}-\frac{H_j W_j^2}{Q^2_j}+(\nu-1)
\Lambda +P_{n-2}[-x_j^2] D_j\right\}\Psi=0\,.\nonumber
\eea
Functions $(D_r,D_j,W_r,W_j)$ and constants $(\Omega,{\tilde\Omega},\Lambda)$ are defined by (\ref{AAdef}). 

When the rotation parameter $a_n$ is sent to zero with fixed $y_n$ (see the discussion of this point above), one gets a degenerate form of the ansatz (\ref{AnstzMaxwEvenOne}):
\bea\label{AnstzMaxwEven}
\hskip -0.5cm
l_{\pm}^\mu A_{\mu}=\left[\frac{r}{r\mp i\mu}+\zeta\right]{\hat l}_\pm \Psi,\ \ 
m_{\pm}^{(k)\mu}  A_{\mu}=\left[\frac{x_k}{x_k\pm \mu}+\zeta\right]{\hat m}^{(k)}_\pm{\Psi}\,,\ \ 
m_{\pm}^{(n)\mu}  A_{\mu}=\zeta{\hat m}^{(n)}_\pm{\Psi}\,.
\eea
Now index $k$ takes values between one and $(n-1)$. In deriving the equation (\ref{MaxwDefrm}) for dimensionally reduced gravitational waves we made two assumptions: the condition $A_\psi=0$ and the absence of $\psi$ dependence in the remaining components of the gauge field\footnote{Recall that in the present case, reduction is performed along $\phi_n$, which should be identified with $\psi$.}. The last relation in (\ref{AnstzMaxwEven}) shows that within our ansatz one of these assumption implies the other. Furthermore, the component $A_{y_n}$ can be eliminated by choosing the gauge $\zeta=0$, and this fact will become very important in section \ref{SecSubProca} when reduction on spheres will be studied. For now we will keep $\zeta$ arbitrary.

Interestingly, for the ansatz  (\ref{AnstzMaxwEven}) with $a_n=0$, the generalized Maxwell's equations (\ref{MaxwDefrmS}) separate even if the warp factor $S$ is more general than (\ref{SfactEvenNG}). Substituting the gauge field from (\ref{AnstzMaxwEven}) into equations (\ref{MaxwDefrmS}), one finds that consistency requires $S$ to have the form
\bea\label{SfactEvenDG}
S=(r\prod_{k<n} x_k)^\nu S_n(y_n)\,,
\eea
where $S_n$ is an {\it arbitrary function} of its argument. The $a_n\rightarrow 0$ limit in the radial and $j=\{1,\dots,n-1\}$ equations in (\ref{EvenDimGenSpin}) is straightforward, but the equation with $j=n$ is modified, so the function $\Psi$ from (\ref{AnstzMaxwEven}) satisfies the following system of ODEs
\bea\label{EvenDimGenRed}
&&\hskip -1cm \frac{D_r}{r^\nu}\frac{\d}{\d r}\left[\frac{r^\nu Q_r^2\Delta}{D_r}
\frac{\d\Psi}{\d r}
\right]-\left\{\frac{2\Lambda}{D_r}-
\frac{R^2 W_r^2}{Q_r^2\Delta}+(\nu-1)\Lambda +P_{n-2}[r^2]  D_r\right\}\Psi=0.\nn
&&\hskip -1cm \frac{D_j}{x_j^\nu}\frac{\d}{\d x_j}\left[\frac{x_j^\nu Q_j^2 H_j}{D_j}
\frac{\d\Psi}{\d x_j}\right]+\left\{\frac{2\Lambda}{D_j}-\frac{H_j W_j^2}{Q^2_j}+(\nu-1)
\Lambda +P_{n-2}[-x_j^2] D_j\right\}\Psi=0,
\\
&&\hskip -1cm \frac{1}{S_n}\frac{\d}{\d y_n}\left[{S_n H_n}\frac{\d\Psi}{\d y_n}\right]+\left\{-{H_n W_n^2}+(\nu+1)
\Lambda +P_{n-2}[0]\right\}\Psi=0\,.\nonumber
\eea
Here
\bea
H_n=(1-y_n^2)\prod_{k<n} a_k^2 \,,\quad
W_n=\omega-\sum_{k<n}\frac{n_k }{a_k}\,,
\eea
and all other ingredients are obtained by taking the $a_n\rightarrow 0$ limit in (\ref{AAdef}).

\bigskip

After this general discussion of separability in equation (\ref{MaxwDefrmS}), we can apply the results to the specific case $S=f^2$ relevant for gravitational waves. Recalling the expression (\ref{FormFactEven}) for function $f$, we conclude that the ansatz (\ref{AnstzMaxwEven}) for gravitational waves leads to the separable system (\ref{EvenDimGenRed}) with 
\bea\label{EvenNuParam}
S=(r\prod_{k<n} x_k)^\nu S_n(y_n)\,,\quad \nu=2,\quad S_n=1-y_n^2\,.
\eea
For a given exponential factor in (\ref{PsiSeparDef}), the free parameters in the system (\ref{EvenDimGenSpin}) are  $\mu$ and $(n-1)$ coefficients of $P_{n-2}$. As expected, this leads to $n$ arbitrary separation constants. The residual gauge invariance parameterized by shifts of parameter $\zeta$ in (\ref{AnstzMaxwEven}) ensures that $\zeta$ does not appear in (\ref{EvenDimGenSpin}). 

\bigskip

To summarize, in this subsection we analyzed separation of variables in the generalized Maxwell's 
equations (\ref{MaxwDefrmS}) in odd--dimensional GLPP geometries. We encountered three cases:
\begin{enumerate}[(a)]
\item If all rotation parameters are nontrivial, then consistency of the separable ansatz (\ref{AnstzMaxwEvenOne}) implies that function $S$ must have the form (\ref{SfactEvenNG}) with an arbitrary parameter $\nu$. The resulting system of ODEs is given by (\ref{EvenDimGenSpin}). For $\nu=0$, this system reduces to separable equations for the standard Maxwell fields studied in \cite{LMaxw}.
\item If one of the rotation parameters vanishes (without loss of generality, we assumed that it is $a_n$), then the separable ansatz is (\ref{AnstzMaxwEven}), the most general consistent warp factor is (\ref{SfactEvenDG}), and the resulting system of ODEs is (\ref{EvenDimGenRed}). 
\item The equation (\ref{MaxwDefrm}) encountered for dimensionally reduced gravitational waves corresponds to a particular case of option (b) with function $S$ given by (\ref{EvenNuParam}).  The relevant system of ODEs is still given by  (\ref{EvenDimGenRed}).
\end{enumerate} 
We also saw that in even dimensions, GLPP geometry with one vanishing rotation parameter has a decoupled two--dimensional sphere, not just $S^1$. So far we have not utilized this enhancement of symmetry, which raises a natural question about extending the reduction (\ref{VectPertDef})--(\ref{VectPertDefAct}) from circles to spheres. We will address this question in the next subsection.

\subsection{Vector modes on spheres and a generalized Proca equation}
\label{SecSubProca}

In the last two subsections we studied reduction of the Einstein--Hilbert action on a circle, which led to decomposition of gravitational waves into scalar, vector, and tensor modes on the base, and we demonstrated full separability of vector modes for the MP and GLPP geometries. In that discussion we implicitly assumed that only one out of ${\bar n}$ rotation parameters vanishes. Other rotation parameters can be send to zero afterwards, and our separation procedure still works. However, if more than one rotation parameter is sent to zero, the geometries contain decoupled spheres rather than circles, and it is possible to find more general modes of gravitational waves which describe polarizations beyond those found in sections \ref{SecVecOdd} and \ref{SecVecEven}. Such polarizations corresponds to general vector modes on spheres reviewed in section \ref{SecModes}, and in this subsection we will construct such modes for degenerate MP and GLPP geometries with two or more vanishing rotation parameters. 

\bigskip

Starting with the MP or  GLPP geometry and setting $p$ out of ${\bar n}$ rotation parameters to zero, one encounters a spacetime with a $S^{2p}$ or $S^{2p-1}$ factors in even or odd dimensions. For the Myers--Perry black hole, this can be easily seen from the metrics  (\ref{MPeven}) and (\ref{MPodd}). Indeed, if one sets 
\bea\label{AzeroOct7}
a_n=\dots a_{n-p+1}=0
\eea
in (\ref{MPeven}), then $2p$ coordinates $(\phi_n,\dots,\phi_{n-p+1},\mu_n,\dots,\mu_{n-p+1})$ appear in the first line of the metric only in the combination
\bea\label{AzeroOct7E2}
\sum_{i=n-p+1}^n r^2 (d\mu_i^2+\mu_i^2 d\phi_i^2)=r^2\mu^2 d\Omega_{2p-1}^2+r^2d\mu^2,
\eea
where
\bea
\mu^2=\sum_{i=n-p+1}^n \mu_i^2\,.\nonumber
\eea
Writing the second line of (\ref{MPeven}) as
\bea
r^2 d\alpha^2=r^2\frac{(\mu d\mu+z dz)^2}{(1-\mu^2-z^2)},\quad z\equiv \left[\sum_{k=1}^{n-p}\mu_k^2\right]^{\frac{1}{2}},\nonumber
\eea
and trading $\mu$ for a new coordinate $\rho$ defined by $\mu=\sqrt{1-z^2}\rho$, we conclude that the cross terms $d\rho dz$ disappear from the metric (\ref{MPeven}), and all contributions involving the sphere $S^{2p-1}$ and $d\rho$ are contained in
\bea
r^2\mu^2 d\Omega_{2p-1}^2+r^2d\mu^2+r^2 d\alpha^2=r^2(1-z^2)\left[
\frac{d\rho^2}{1-\rho^2}+\rho^2 d\Omega_{2p-1}^2\right]+\frac{r^2 z^2dz^2}{1-z^2}
\eea
The terms in the square brackets describe the metric on a unit $S^{2p}$. Therefore, starting with the MP geometry in even dimensions and setting $p$ out of $\bar n$ rotation parameters to zero, one arrives at the metric 
\bea\label{MPresSn}
ds^2&=&-dt^2+\frac{Mr}{FR}\Big(dt+\sum_{i=1}^q a_i\mu_i^2 d\phi_i\Big)^2+\frac{FR dr^2}{R-Mr}
+\sum_{i=1}^q(r^2+a_i^2)\Big(d\mu_i^2+\mu_i^2 d\phi_i^2\Big)\nn
&&+\frac{r^2 z^2dz^2}{1-z^2}+r^2(1-z^2) d\Omega_{2p}^2\,,\nonumber
\eea
which has the form (\ref{SpaceSplit}) with $s=2p$ and $f^2=r^2(1-z^2)$. 
 
Similarly, in the odd--dimensional case, the constraint (\ref{AzeroOct7}) ensured that the coordinates $(\phi_{\bar n},\dots,\phi_{{\bar n}-p+1},\mu_{\bar n},\dots,\mu_{{\bar n}-p+1})$ appear in the metric (\ref{MPodd}) only through combinations (\ref{AzeroOct7E2}), so the $S^{2p-1}$ factor is present in this case as well, but this time there is no enhancement to $S^{2p}$. Therefore, starting with the MP geometry in even dimensions and setting $p$ out of $n$ rotation parameters to zero, one arrives at the metric (\ref{SpaceSplit}) with 
$s=2p-1$ and with $f=r\mu$. Similar arguments can be used to justify enhanced symmetries for of the GLPP black holes in the presence of the constraint (\ref{AzeroOct7}). Note that regardless of the starting point, the dimension of the base metric ${\bar g}_{\mu\nu}$ always turns out to be even\footnote{Starting with $d=2(n+1)$ and removing $S^{2p}$, one gets $2(n-p+1)$ dimensions. Starting with $d=2n+3$ and removing $S^{2p-1}$, one arrives at a $2(n-p+2)$--dimensional base.}, and the special frames on this base are given by setting 
$a_{\bar n}=\dots a_{{\bar n}-p+1}=0$ in
\bea\label{ProcaFrameBase}
m^{(0)}_\pm,\quad m^{(j)}_\pm,\quad j=1,\dots,q,\quad q=n-p+d-2\left[\frac{d}{2}\right]\,.
\eea
In other words, $q$ is equal to $(n-p)$ in even and $(n-p+1)$ in odd dimensions. Recall that for nonzero values of rotation parameters, the frames in even and odd dimensions are given by (\ref{Mframes}) and (\ref{GenFramesOddD}). 

\bigskip

Once the presence of the $S^{s}$ factor is established, the metric is guaranteed to take the form (\ref{SpaceSplit}). Then one can start with the most general vector perturbation (\ref{ClasVec1}) and perform a gauge transformation
\bea
y_a\rightarrow y^a-h_V V^a
\eea
to put the excitation in the form similar to (\ref{PertVec1})
\bea\label{SphereVec}
ds^2_\eps=2\eps V_a h^{(V)}_\mu dx^\mu dy^a,\quad h^{(V)}_\mu=f^2 A_\mu\,.
\eea
Alternatively, one can keep both terms in (\ref{ClasVec1}) and formulate the problem in terms of a gauge invariant quantity
\bea
F_\mu=\frac{1}{f}h_\mu+\frac{f}{2k}\d_\mu \frac{h_V}{f^2}\,.
\eea
Articles \cite{KodamaOne,5DGravWaveW1,5DGravWaveW2,5DGravWaveW3,5DGravWaveW4,5DGravWaveW5,
5DGravWaveW6} chose the latter gauge--invariant option, but to avoid unnecessary complications, we will focus on perturbation (\ref{SphereVec}) without losing generality. Recalling that the vector modes $V_a$ satisfy differential equations (\ref{ClasVec2}) with the eigenvalues (\ref{SphVecEigenOne}),
\bea\label{ClasVec2a}
({\hat \nabla}^2+\la_V) V_a=0,\quad {\hat \nabla}_a V^a=0,\quad \la_V=\ell (\ell+s-1)-1,\quad \ell=1,2,\dots
\eea
one finds \cite{KodamaOne} that the linearized Einstein's equations for the perturbation (\ref{SphereVec}) reduces 
to 
\bea
\frac{1}{f^{s+1}}{\bar\nabla}^\mu\left[f^{s+2}F_{\mu\nu}\right]-\frac{\la_V-(s-1)}{f}A_\nu=0,
\quad
{\bar\nabla}_\mu(f^{s}A^\mu)=0.
\eea
Here the covariant derivatives are taken on the $(d-s)$--dimensional base of the geometry (\ref{SpaceSplit}) with metric ${\bar g}_{\mu\nu}$. Using covariant derivatives with respect to the entire $d$--dimensional space instead, one finds
\bea\label{SnReducMN}
\frac{1}{f^2}\nabla^M\left[f^{2}F_{MN}\right]-\frac{\la_V-(s-1)}{f^2}A_N=0,
\quad
\nabla_M A^M=0.
\eea
The components of the first equation along the sphere directions are trivially satisfied since we assumed that the gauge field had the form $A=A_\mu (x)dx^\mu$. Relations (\ref{SnReducMN}) can be interpreted as a generalized Proca equation with an additional constraint. 

\bigskip

Motivated by (\ref{SnReducMN}), in the remaining part of this subsection we will study a generalized Proca equation of the form
\bea\label{GenProca}
\frac{1}{S}\nabla_\mu\left[S F^{\mu\nu}\right]-M^2 S^\sigma A^\nu=0,
\eea
where $S$ is some undetermined function, and $\sigma$ is a free parameter. The background geometry is obtained by setting $p$ rotational parameters $a_k$ to zero, and the metric of the base space has reduced frames (\ref{ProcaFrameBase}). Using inspiration from the previous two subsections and from the study of the Proca equation carried out in \cite{Proca}, we impose an ansatz on the vector field
\bea\label{AnstzMaxwProca}
m_{\pm}^{(0)\mu} A_{\mu}=\left[\frac{r}{r\mp i\mu}+\zeta\right]{\hat m}^{(0)}_\pm \Psi,\quad
m_{\pm}^{(j)\mu}  A_{\mu}=\left[\frac{x_j}{x_j\pm \mu}+\zeta\right]{\hat m}^{(j)}_\pm{\Psi},\quad
j=1\dots q\,
\eea
with a separable function $\Psi$ given by
\bea\label{PsiSeparDefProca}
\Psi=E\Phi(r)\left[\prod_{j}^q X_j(x_j)\right],\qquad E=\exp\left[i\omega t+i\sum_j^q n_j\phi_j\right]\,.
\eea
Note that in the presence of the mass term for the vector, there is no gauge invariance, so parameter $\zeta$ is fixed by equations of motion, as we will see below. Since equation (\ref{GenProca}) must be satisfied for all values of $M$, including $M=0$, the arguments which led to (\ref{SfactOdd}) and  (\ref{SfactEvenNG}) imply that the  factor $S$ must have the form
\bea\label{SfactNDproc}
S=(r\prod_k^q x_k)^\nu\,
\eea
with a free parameter $\nu$. Substituting the resulting vector field into equation (\ref{AnstzMaxwProca}) with a {\it nonzero} value of $M$, one finds only two consistent options:
\label{LabBrnchAB}
\begin{enumerate}[a),]
\item \underline{$\zeta=0$} \\
In this case, parameter $\sigma$ is also fixed, and the vector field satisfies a constraint:
\bea
\sigma=-\frac{2}{\nu},\quad 
\nabla_\mu[S^{1+\sigma} A^\mu]=0\,.
\eea
The generalized Proca equation (\ref{GenProca}) reads
\bea\label{GenProcaZeta}
\frac{1}{S}\nabla_\la\left[S F^{\la\mu}\right]-\frac{M^2}{S^{2/\nu}} A^\mu=0,
\eea
In particular, for $\nu=2$, this branch recovers equations (\ref{SnReducMN}) for the vector modes of gravitational waves on a sphere. This option leads to consistent separable equations if and only if at least one rotation parameter vanishes. Equations (\ref{SnReducMN}) were obtained under this assumption as well.
\item \underline{$\zeta=-1$}\\
In this case, the value of $\sigma$ is fixed to be $\sigma=0$, and equations for the vector field are 
\bea\label{GenProcaZetaM}
\frac{1}{S}\nabla_\mu\left[S F^{\mu\nu}\right]-M^2 A^\nu=0,\quad 
\nabla_\mu[S A^\mu]=0.
\eea
In particular, for $\nu=0$, or $S=1$, one recovers the separable solutions of the standard Proca equation constructed in \cite{Proca}. Unless $\nu=0$, this branch gives consistent separable solutions if and only if at least one rotation parameter vanishes. The solutions with $\nu=0$ work even for the non--degenerate GLPP geometry.
\end{enumerate}
To complete the discussion of the generalized Proca equation (\ref{GenProca}) in the (degenerate) MP and GLPP backgrounds, let us present the differential equations  for various ingredients appearing in (\ref{PsiSeparDefProca}). Not surprisingly, equations look somewhat different in even and odd 
dimensions\footnote{Here dimension refers to $d$, not to the dimension of the base, which is always even.}. 

\bigskip
\noindent
{\bf Even dimensions:}\\
The differential equations read
\bea\label{EvenDimProca}
&&\hskip -1cm
\frac{D_r}{S}\frac{\d}{\d r}\left[\frac{S Q_r^2\Delta}{D_r}\frac{\d\Psi}{\d r}\right]-\left\{\frac{2\Lambda}{D_r}-
\frac{R^2 W_r^2}{Q_r^2\Delta}+(\nu-1)\Lambda +(-r^2)^{s/2}P_{q-2}[-r^2]  D_r+M^2G_r\right\}\Psi=0,
\nn
\ \\
&&\hskip -1cm
\frac{D_j}{S}\frac{\d}{\d x_j}\left[\frac{S Q_j^2 H_j}{D_j}\frac{\d\Psi}{\d x_j}\right]+\left\{\frac{2\Lambda}{D_j}-\frac{H_j W_j^2}{Q^2_j}+(\nu-1)
\Lambda +(x_j)^sP_{q-2}[x_j^2] D_j+M^2G_j\right\}\Psi=0\,.\nonumber
\eea
Here functions $(G_r,G_j)$ defined by
\bea\label{ProcaGfact}
\zeta=0:&&G_r=\frac{D_r}{r^2},\quad G_j=\frac{D_j}{(x_j)^2}\,;\nn
\zeta=-1:&&G_r=\mu^2 D_r r^{2(n-1)},\quad G_j=\mu^2 D_j (-x_j^2)^{n-1}\,.
\eea
As already mentioned, at least one rotation parameter must be set to zero, unless one chooses the second option and sets $\nu=0$. 

\bigskip
\noindent
{\bf Odd dimensions:}\\
The differential equations read
\bea\label{OddDimProca}
&&\hskip -1cm
\frac{D_r}{rS}\frac{\d}{\d r}\left[\frac{S Q_r^2\Delta}{rD_r}\frac{\d\Psi}{\d r}\right]+
\left\{\frac{2\Lambda}{D_r}+
\frac{R^2 W_r^2}{r^2Q_r^2\Delta}+\nu\Lambda +(-r^2)^{\frac{s-1}{2}}P_{q-2}[-r^2]  D_r-M^2G_r\right\}\Psi=0,
\nn
\ \\
&&\hskip -1cm
\frac{D_j}{x_jS}\frac{\d}{\d x_j}\left[\frac{S Q_j^2 H_j}{x_jD_j}\frac{\d\Psi}{\d x_j}\right]+\left\{\frac{2\Lambda}{D_j}-\frac{H_j W_j^2}{x_j^2Q^2_j}+\nu
\Lambda +(x_j)^{s-1}P_{q-2}[x_j^2] D_j+M^2G_j\right\}\Psi=0\,.\nonumber
\eea
The factors $(G_r,G_j)$ are still given by (\ref{ProcaGfact}). While writing equations, we assumed that at least one of the rotational parameters is zero. As already mentioned, in the non--degenerate case, it is only the standard Proca equation (option (\ref{GenProcaZetaM}) with $\nu=0$) that separates with the ansatz (\ref{AnstzMaxwProca}). In this case, equations (\ref{OddDimProca}) contain additional $\AA$--terms, which are identical to those appearing in (\ref{OddMPmaster}). 

\bigskip
\noindent
To summarize, in this subsection we analyzed vector gravitational modes in the geometry (\ref{MPresSn}) and its odd--dimensional counterpart. The excitations have the form (\ref{SphereVec}), and their dynamics is governed by the generalized Proca equation (\ref{SnReducMN}). We demonstrated separability of this equation using the ansatz (\ref{AnstzMaxwProca}), and reduced the problem to systems of ODEs (\ref{EvenDimProca}) and (\ref{OddDimProca}). To recover the specific equation  (\ref{SnReducMN}) rather than its more general version, one should set 
\bea\label{ProcaZetaNu}
\zeta=0,\quad \nu=2,\quad M^2=\la_V-(s-1)=\ell (\ell+s-1)-s
\eea
in the systems  (\ref{EvenDimProca}), (\ref{ProcaGfact}), (\ref{OddDimProca}). The separation constants are $\mu$, $(q+1)$ parameters in (\ref{PsiSeparDefProca}) and $(q-1)$ coefficients of $P_{q-2}$\footnote{For $q=1$, this polynomial is absent: $P_{-1}[x]\equiv 0$.}. This counting reproduces the 
$(2q+1)$ free parameters expected for the most general separation in $2(q+1)$ dimensions. It is clear that the counting should work differently for $q=0$, and given the importance of this special case, we will briefly discuss it in the next subsection.

\subsection{The static limit}
\label{SecSubStatic}

In this subsection we consider the vector mode of gravitational waves in the Schwarzschild--Tangherlini geometry \cite{Tang}
\bea
ds^2=-h dt^2+\frac{dr^2}{h}+r^2 d\Omega_s^2
=g_{\mu\nu}dx^\mu dx^\nu+r^2 d\Omega_s^2,\quad h=1-\frac{2m}{r^{s-1}}
\eea
This problem has been extensively studied in the past using gauge--invariant variables \cite{KodamaOne,5DGravWaveW1,5DGravWaveW2,5DGravWaveW3,5DGravWaveW4,5DGravWaveW5,
5DGravWaveW6}, but to compare to the results obtained in the previous subsection, we need to formulate the answer in terms of the gauge potential. The perturbation is given by (\ref{SphereVec}), and the most general separable gauge field in the $(t,r)$ subspace has the form
\bea\label{AntzSchw}
A_\mu dx^\mu=e^{i\omega t}\left[A_t(r)dt+A_r(r)dr\right],\quad
h_\mu^{(V)}=r^2 A_\mu\,.
\eea
Equations (\ref{SnReducMN}) can be written in terms of the metric on the two--dimensional base as
\bea\label{SnReducSchw}
\frac{1}{r^{2+s}}\d_\la\left[r^{2+s}F^{\la\mu}\right]-\frac{M^2}{f^2}A^\mu=0,
\quad
\d_\mu(r^s A^\mu)=0.
\eea
As before, it is instructive to consider a counterpart of equation (\ref{GenProcaSchw}), which generalizes (\ref{SnReducSchw}) by introducing an arbitrary function $S(r)$ and a parameter $\sigma$:
\bea\label{GenProcaSchw}
\frac{1}{Sr^{s}}\d_\la\left[Sr^{s}F^{\la\mu}\right]-M^2 S^\sigma A^\mu=0.
\eea
Substituting the gauge potential (\ref{AntzSchw}) into these equations, we find the expression for $A_r$,
\bea\label{EqnSchw1}
A_r=\frac{i\omega A'_t}{M^2 hS^\sigma-\omega^2},
\eea
and a differential equation for $A_t$:
\bea\label{EqnSchw2}
\frac{1}{r^{s+1}}
\frac{d}{dr}\left[\frac{r^{s+1} h S^{\sigma+1}A'_t}{M^2 hS^\sigma-\omega^2}\right]-
\frac{S^{\sigma+1}}{h}A_t=0
\eea
This equation ensures that the gauge potential satisfies the relation
\bea
\d_\la(r^s S^{\sigma+1} A^\la)=0.
\eea
Equations (\ref{SnReducSchw}) are reproduced if one sets 
\bea\label{SfuncSchw}
S=r^2,\quad \sigma=-1.
\eea
The same expressions were encountered in the last subsection. For nonzero $\omega$, equations (\ref{EqnSchw1})--(\ref{EqnSchw2}) can be rewritten as
\bea
&&A_t=
\frac{h}{T}\frac{d}{dr}\left[\frac{hT A_r}{i\omega}\right],\quad B_r=hT A_r,\nn
&&{hT}\frac{d}{dr}\left[\frac{h}{T}\frac{dB_r}{dr}\right]+
(\omega^2-M^2 hS^\sigma)B_r=0.
\eea
Here $T\equiv S^{\sigma+1}r^{s+1}$. The structure of the differential equation for $B_r$ is similar to that of equations (\ref{EvenDimProca}), (\ref{OddDimProca}), although this time there are no separation constants. Solutions of the differential equation for $B_r$ with parameters (\ref{SfuncSchw}) lead to the most general vector modes of gravitational waves described by equations (\ref{SphereVec}) and (\ref{SnReducSchw}).

\section[An example of a vector mode of a general rotating black hole]{\large An example 
of a vector 
mode of a general rotating black hole}
\label{SecHair}

In the last section we analyzed gravitational waves in rotating black holes with at least one rotation parameter set to zero. In this case, one can apply the classification presented in section \ref{SecModes}, and we focused on vector modes. When all rotation parameters are nontrivial, the split into scalar, vector, and tensor modes does not apply, but in this section we will argue that in odd dimensions, it might be possible to isolate certain polarizations similar to the vector modes discussed earlier.  We will call such polarizations ``vector modes'', and the goal of this section is to construct some special cases of such excitations and discuss challenges associated with adding more quantum numbers to these modes. To avoid unnecessary complications, we will mostly focus on the five--dimensional case and set the cosmological constant to zero.

\bigskip

The expression (\ref{AnstzMaxw}) for the vector modes of gravitons in odd dimensions is very suggestive, and one may hope to extend this ansatz to geometries with arbitrary non--zero rotation parameters. One such mode corresponds to a ``black hole hair'' obtained by varying parameters of the solution, and in this section we will use inspiration from this hair to construct more general solutions covered by extensions of the ansatz (\ref{AnstzMaxw}). We begin with observing that a perturbation
\bea\label{PertWithA}
h_{\mu\nu}=\sum A_i\frac{\d}{\d a_i} g_{\mu\nu}^{(background)}
\eea
solves linearized Einstein's equations for arbitrary constants $A_i$. This perturbation describes a difference between two MP black holes with rotation parameters $a_k$ and 
\bea
{\tilde a}_k=a_k+\eps A_k\,.
\eea
In particular, we will focus on 
\bea\label{OperAnn}
A_k=\frac{c}{a_k}
\eea
for some constant $c$. If the original solution has $a_n=0$, then 
\bea
\sum A_i\frac{\d}{\d a_i}\propto \frac{\d}{\d a_n}\,,
\eea
and perturbations have the form
\bea
h_{\mu\nu}dx^\mu dx^\nu=2h_{\mu\psi}dx^\mu d\psi\,.
\eea
Such modes were discussed in the previous section, and now we will analyze application of the operator (\ref{OperAnn}) to a general MP solution in odd dimensions. To avoid unnecessary complications, we will focus on $d=5$ and discuss extensions to other dimensions in the end of this section.

\bigskip

For our discussion, it is convenient to write the Myers--Perry geometry in terms of a new coordinate $\theta$ defined by the relation $\mu_1=\sin\theta$. In five dimensions, there is only one $x$ coordinate, and it is 
given by\footnote{In this section we use shorthand notation $s_\theta=\sin\theta$, $c_\theta=\cos\theta$.}
\bea
x_1=\sqrt{(ac_\theta)^2+(bs_\theta)^2}\equiv \Theta
\eea
The inverse metric van be written as
\bea\label{Metr5D}
g^{\mu\nu}\d^\mu\d^\nu=\frac{1}{\Sigma\Delta}l_+^\mu l_-^\nu\d_\mu\d_\nu+
\frac{1}{\Sigma}m_+^\mu m_-^\nu\d_\mu\d_\nu+\frac{1}{r\Theta}n^\mu n^\nu\d_\mu\d_\nu\,,
\eea
where function $\Sigma$ is defined by
\bea
\Sigma=r^2+\Theta^2\,.
\eea
Introducing a new coordinate $q=c_\theta$, we can write the reduced frames as
\bea
\label{LCframes5D}\label{AllFramesMPOdd5D}
l^\mu_\pm&=&\sqrt{\Sigma\Delta}\,(e^\mu_r\pm e^\mu_t),\quad 
m^\mu_\pm=\sqrt{\Sigma}\,(e^\mu_q\pm i e^\mu_1),\quad 
n^\mu=r\Theta\, e_\psi^\mu\,,\nn
e_t&=&\frac{R}{r\sqrt{\Sigma\Delta}}\left[ \d_t
-\sum_k\frac{a_k}{r^2+a_k^2}\d_{\phi_k}\right],\quad 
e_r=\sqrt{\frac{\Delta}{r^2\Sigma}}\d_r,
\quad 
e_{q}=\sqrt{\frac{1-q^2}{\Sigma}}\d_{q},\\
e_1&=&\frac{q\sqrt{1-q^2}}{\Theta\sqrt{\Sigma}}\left[(a^2-b^2)\d_t+\frac{a}{1-q^2}\d_\phi
-\frac{b}{q^2}\d_\psi
\right],\
e_\psi=\frac{1}{r\Theta}\left[ab\d_t+b\d_\phi+a\d_\psi
\right].\nonumber
\eea
For vanishing mass, equation (\ref{Metr5D}) describes the metric of flat space in ellipsoidal coordinates regardless of the values of $(a,b)$, therefore perturbation (\ref{PertWithA}) can be removed by performing a diffeomorphism. To isolate the nontrivial part of the perturbation, we consider
\bea\label{h5ddif}
h_{\mu\nu}=\frac{1}{2m}(b\d_a+a\d_b)g_{\mu\nu}+
\frac{1}{m}(\nabla_\mu \xi_\nu+\nabla_\nu \xi_\mu)
\eea
and choose the generator of the diffeomorphism to be 
\bea
\xi^\mu\d_\mu=-\frac{ab}{2r}\d_r\,.
\eea
With this choice, inverse powers of mass disappear, and the perturbation becomes 
\bea
h_{\mu\nu}dx^\mu dx^\nu&=&\frac{2}{\Sigma}[bp^2 d\phi^2+aq^2 d\psi^2]dt
-\frac{2(a^2+b^2)(pq)^2}{\Sigma}d\phi d\psi\nn
&&-
\frac{2ab}{\Sigma}[p^4 d\phi^2+q^4 d\psi^2]+\frac{2abr^2\Sigma}{\Delta^2}dr^2
\eea
The projections of this perturbation on the special frames (\ref{LCframes5D}) look remarkably simple,
\bea\label{HairProjDif}
&&
n^\mu l_\pm^\nu h_{\mu\nu}=\pm \frac{r^2\Theta^2}{\Delta}l^r_+,\quad l^\mu_+l^\nu_+h_{\mu\nu}=l^\mu_-l^\nu_-h_{\mu\nu}=\frac{4abr^2\Sigma}{\Delta^2}(l_+^r)^2,\\ 
&&n^\mu n^\nu h_{\mu\nu}=0,\quad m_\pm^\mu h_{\mu\nu}=0,\quad
l^\mu_+l^\nu_-h_{\mu\nu}=0,\nonumber
\eea
and we will use them as an inspiration for extending the perturbation  beyond the ``black hole hair'' (\ref{h5ddif}). The extensions will go in two steps:
\begin{enumerate}
\item Keeping the second line of (\ref{HairProjDif}) intact, generalizing the first line as
\bea
n^\mu l_\pm^\nu h_{\mu\nu}=\Theta^2 f^{(1)}_\pm(r),\quad 
l^\mu_\pm l^\nu_\pm h_{\mu\nu}=\Sigma f^{(2)}_\pm(r)
\eea
with four undetermined functions $(f^{(1)}_\pm,f^{(2)}_\pm)$, and substituting the result into linearized Einstein's equations, we arrive at the most general solution of this type:
\bea\label{SeparRR}
&&n^\mu l_\pm^\nu h_{\mu\nu}=\frac{r^2\Theta^2}{\Delta}(f_1\pm c_1)l^r_+,\quad 
l^\mu_\pm l^\nu_\pm h_{\mu\nu}=\frac{4abr^2\Sigma}{\Delta^2}[c_1+c_2 r^2\pm f_2](l_+^r)^2\,,
\nn
&&n^\mu n^\nu h_{\mu\nu}=0,\quad m_\pm^\mu h_{\mu\nu}=0,\quad
l^\mu_+l^\nu_-h_{\mu\nu}=0.
\eea
Here $(c_1,c_2)$ are free parameters, and $(f_1,f_2)$ are arbitrary functions of the radial coordinate which correspond to pure gauges. Solution (\ref{HairProjDif}) is recovered for $f_1=f_2=c_2=0$.

It is instructive to compare this perturbation with $b=0$ to the general solution (\ref{AnstzMaxw}) with a vanishing rotation parameter. To do so, we assume that $\Psi$ in (\ref{AnstzMaxw}) does not depend on cyclic coordinates. Then equations (\ref{AnstzMaxw}) and (\ref{ODEoddMgen}) can be rewritten in terms of the $(r,q)$ variables as
\bea\label{EqnOne}
&&l_{\pm}^\mu n^\nu h_{\mu\nu}=q^2 r^2\left[\frac{r}{r\mp \mu}+\zeta\right]{\hat l}_+ \Psi,\quad
m_{\pm}^\mu n^\nu h_{\mu\nu}=q^2r^2\left[\frac{aq}{aq\pm i\mu}+\zeta\right]{\hat m}_+{\Psi}\nn
&&\d_r\left[\frac{r\Delta}{r^2-\mu^2}\d_r\Psi\right]=0,\quad \d_q\left[\frac{q^3(1-q^2)}{\mu^2+(aq)^2}\d_q\Psi\right]=0.
\eea
In particular, the radial derivative of function $\Psi$ is
\bea
\d_r\Psi=c\frac{r^2-\mu^2}{r\Delta}
\eea
with a free integration constant $c$. Substitution into the first line of (\ref{EqnOne}) gives
\bea
l_{\pm}^\mu n^\nu h_{\mu\nu}=\frac{r^3 q^2}{r\mp \mu}l^r_+\d_r \Psi=
\frac{c r^2 q^2(r\pm \mu)}{\Delta}l^r_+
\eea
This expression matches (\ref{SeparRR}) with $f_1=\frac{c_1 r}{\mu}$. Projections 
$l^\mu_\pm l^\nu_\pm h_{\mu\nu}$ in (\ref{SeparRR}) vanish since $b=0$.
\item
Inspired by (\ref{EqnOne}), one can generalize (\ref{SeparRR}) to 
\bea\label{EqnTwo}
&&n^\mu l_\pm^\nu h_{\mu\nu}=\pm {c_1\Theta^2 r}(l_+^r)\d_r\Psi,\quad 
l^\mu_\pm l^\nu_\pm h_{\mu\nu}=\frac{4ab\Sigma r(c_1+c_2 r^2)}{\Delta}
(l_+^r)^2\d_r\Psi,\nn
&& l^\mu_+ l^\mu_- h_{\mu\nu}=0,\quad  m^\mu_+ m^\mu_- h_{\mu\nu}=0,\quad
n^\mu n^\nu h_{\mu\nu}=0,\\
&&n^\mu m^\nu_\pm h_{\mu\nu}=\mp ic_1\Theta r^2(m_+^q)\d_q\Psi,\
m^\mu_\pm m^\nu_\pm h_{\mu\nu}=-\frac{4abq\Sigma(c_1-c_2\Theta^2)}{(pq)^2(a^2-b^2)}(m_+^q)^2\d_q\Psi,
\nonumber
\eea
where function $\Psi$ satisfies a system of decoupled differential equations
\bea\label{EqnTwoA}
\frac{\d}{\d r}\left[\frac{\Delta}{r}\frac{\d\Psi}{\d r}\right]=0,\quad \frac{\d}{\d q}\left[\frac{q^2(1-q^2)}{q}\frac{\d \Psi}{\d q}\right]=0
\eea
The solution (\ref{SeparRR}) with $f_1=f_2=0$ is recovered for $q$--independent function $\Psi$. 
For $b=0$, the solution (\ref{EqnTwo})--(\ref{EqnTwoA}) reproduces the system (\ref{EqnOne}) with 
$\mu=0$.  Unfortunately, we were not able to extend the system  (\ref{EqnTwo})--(\ref{EqnTwoA}) to equations with a separation constant similar to $\mu$ in (\ref{EqnOne})  unless $ab=0$, and this also obstructs an inclusion of dependence on cyclic coordinates. 

\end{enumerate}
Extensions of the solutions found in this section to other odd dimensions is straightforward. For example, the ``black hole hair'' described by (\ref{PertWithA}) and (\ref{OperAnn}) with a compensating diffeomorphism reads
\bea\label{HairProjDifDD}
&&\hskip -1cm
n^\mu l_\pm^\nu h_{\mu\nu}=\pm \frac{r^2\prod x_j^2}{\Delta}l^r_+,\quad l^\mu_+l^\nu_+h_{\mu\nu}=l^\mu_-l^\nu_-h_{\mu\nu}=\frac{4r^2}{\Delta^2}\left[\prod a_j(r^2+a_j^2)\right](l_+^r)^2,\\ 
&&\hskip -1cm n^\mu n^\nu h_{\mu\nu}=0,\quad m_\pm^\mu h_{\mu\nu}=0,\quad
l^\mu_+l^\nu_-h_{\mu\nu}=0,\nonumber
\eea
The extension of (\ref{EqnTwo})--(\ref{EqnTwoA}) follows the same pattern, and it still depends on two constants $(c_1,c_2)$.

\bigskip
\noindent
To summarize, in this section we analyzed the gravitational perturbations obtained by varying the Myers--Perry solutions with respect to rotation parameters. We used the result as an inspiration for constructing a more general solution (\ref{EqnTwo})--(\ref{EqnTwoA}) and its extensions to higher dimensions. If one of the rotation parameters is set to zero, this solution matches the perturbation (\ref{AnstzMaxwGW}) discussed in the previous section, but only for $\mu=0$ and only for $\Psi(r,x_1\dots x_n)$. By generalizing the solution (\ref{EqnTwo})--(\ref{EqnTwoA}) to arbitrary values of $\mu$ and by adding dependence on cyclic coordinates, one would be able to separate equations for gravitational waves on the MP background with arbitrary rotation parameters, and we hope to address this interesting generalization in the future.

\section{Scalar modes of partially rotated black holes}
\label{SecScalar}

In section \ref{SecVector} we analyzed black holes with $({\bar n}-1)$ rotation parameters and constructed vector modes of gravitational waves. In this section we will briefly discuss scalar modes of such black holes, as well as some excitations of geometries with less than  
$({\bar n}-1)$ rotations.

\subsection{Reduction on a circle}
\label{SecSubScalCrcl}

Let us go back to the background geometry (\ref{MetrFact}) and focus on $\psi$--independent gravitational waves. Due to symmetry under reflection of $\psi$, excitations split into decoupled odd and even modes, and vector perturbation given by (\ref{VectPertDef}) constitute the most general solutions in the former case. For the even excitations, the most general metric has the form
\bea\label{PertScalStart}
ds^2=g_{\mu\nu}dx^\mu dx^\nu+f^2 d\psi^2+
\eps\left[h_{\mu\nu}dx^\mu dx^\nu+h_{\psi\psi} d\psi^2\right]
\eea
where $(h_{\mu\nu},h_{\psi\psi} )$ are functions of coordinates $x^\mu$. This leads to two distinct options:
\begin{enumerate}[(a)]
\item If $f$ is a nontrivial function, then the mode $h_{\psi\psi}$ can be removed by making a diffeomorphism which preserves the form of (\ref{PertScalStart}). For example, if $f=x^1$, then the relevant diffeomorphism is
\bea
x^1\rightarrow x^1-\eps \frac{h_{\psi\psi}}{2x^1}
\eea
For preservation of the form of (\ref{PertScalStart}), specifically for the absence of cross terms $dx^\mu d\psi$, it is crucial that $\d_\psi h_{\psi\psi}=0$. Therefore, for a nontrivial function 
$f$, one can set $h_{\psi\psi}=0$ and focus only in the tensor modes $h_{\mu\nu}$. Construction of such excitations is beyond the scope of this article.
\item If $f$ is a constant, then $h_{\psi\psi}$ cannot be removed by a coordinate transformation, and this scalar mode leads to a dynamical gravitational wave. Such situation will be analyzed in the rest of this subsection.
\end{enumerate}
Based on the comments above, we will focus on the background metric and excitations given by
\bea\label{PertScalStartV2}
ds^2=g_{\mu\nu}dx^\mu dx^\nu+d\psi^2+
\eps\left[h_{\mu\nu}dx^\mu dx^\nu+h_{\psi\psi} d\psi^2\right]
\eea
and look for solutions with a nontrivial $h_{\psi\psi}$. The simplest option for such excitations is obtained by requiring $h_{\mu\nu}$ to be proportional to the background metric. In other words, we will focus on scalar excitations in the form
\bea\label{PertScalStartV3}
ds^2=(1+\eps \Phi){g}_{\mu\nu}dx^\mu dx^\nu+(1+\eps\Psi)d\psi^2,
\eea
where $(\Phi,\Psi)$ are functions of coordinates $x^\mu$. The linearized Einstein's equations give
\bea\label{WaveEqnS1}
\Phi=-\frac{1}{p-2}\Psi,\quad
\frac{1}{\sqrt{-g}}\d_\mu\left[\sqrt{-g}g^{\mu\nu}\d_\nu \Psi\right]=0,
\eea
where $p$ is the dimension of the metric ${g}_{\mu\nu}$.
In the context of the Myers--Perry geometry, the background metric (\ref{PertScalStartV3}) with $\eps=0$ arises when one takes a product of a black hole and a circle. In this case, the metric $g_{\mu\nu}$ on the base is just the MP geometry, and the wave equation (\ref{WaveEqnS1}) is known to be separable. The details of this separation are presented in the Appendix \ref{SecMPwave}. 

An extension of (\ref{PertScalStartV3}) to several compact variables is straightforward:
\bea\label{PertScalStartV4}
ds^2=\left(1-\eps\sum_k \frac{\Psi_k}{p-2}\right)g_{\mu\nu}dx^\mu dx^\nu+\sum_k (1+\eps\Psi_k)d\psi_k^2+
\eps \sum_{i\ne j}\Psi_{ij}d\psi_jd\psi_j,
\eea
Here all functions $\Psi_k$, $\Psi_{ij}$ satisfy the wave equation (\ref{WaveEqnS1}). Note that in the case of several coordinates $\psi_k$, excitations $\Psi_k$ in a counterpart of equation (\ref{PertScalStart}) with nontrivial prefactors, 
\bea\label{PertScalStartV6}
ds^2=[g_{\mu\nu}+\eps h_{\mu\nu}]dx^\mu dx^\nu+\sum f_k^2 (1+\eps \Psi_k)d\psi_k^2
\eea
may or may not be removable by diffeomorphisms, depending on the structure of coefficients 
$f_k$. From the perspective of the MP geometry, this toric reduction (\ref{PertScalStartV6}) with nontrivial functions $f_k$ arises when two or more rotation parameters are set to zero, and in this case coordinates $\psi_k$ along with some non--cyclic directions combine into spheres, as discussed in the beginning of section \ref{SecSubProca}, leading to symmetry enhancement. The scalar modes of gravitational waves emerging from reduction on such spheres, which is more interesting than the toric case, will be discussed in the next subsection. 

\subsection{Reduction on a sphere}
\label{SecSubScalSph}

In this section we are decomposing the space into a $p$--dimensional base and the remaining directions, which can form either a torus or a sphere, and we are classifying the modes according to their transformation properties under diffeomorphisms on the base. This is the most reasonable convention  for reduction on tori discussed in the previous subsection, but in the case of reduction along a sphere $S^s$ one can also use an alternative convention based on transformations of modes under $SO(s+1)$. This convention has been extensively used in the literature \cite{KodamaOne,5DGravWaveW1,5DGravWaveW2,5DGravWaveW3,5DGravWaveW4,5DGravWaveW5,
5DGravWaveW6,KodamaRotOneAw1,KodamaRotOneAw2,AllDimOtherW1,AllDimOtherW2,AllDimOtherW3}, and we reviewed it in section \ref{SecModes}. In particular, scalar modes on the base constitute tensor modes on the sphere, (\ref{TensoModeDef})--(\ref{TensoModeDef2}), since all indices of such excitations point in the sphere direction. The goal of this subsection is to analyze the {\it scalar modes on the base} of degenerate Myers --Perry geometries using Einstein's equations for the modes (\ref{TensoModeDef})--(\ref{TensoModeDef2}) .

\bigskip

Let us consider the metric containing the perturbation (\ref{TensoModeDef}):
\bea\label{TensPertNew}
ds^2={\bar g}_{\mu\nu}dx^\mu dx^\nu+f^2 d\Omega_s^2+\eps f^2 h(x) T_{ab}(y)dy^ady^b\,,
\eea
where the tensor $T_{ab}$ on the $s$--sphere satisfies the relations (\ref{TensoModeDef2}). As reviewed in the Appendix \ref{AppSphHarm}, the eigenvalues of tensor Laplacian on an $s$--sphere are given by
\bea
\lambda_T=l(l+s-1)-2.
\eea
Substituting the metric (\ref{TensPertNew}) into linearized Einstein's equations and using properties of tensor harmonics, one arrives at a single equation for the scalar $h$ \cite{KodamaOne}:
\bea\label{HtensEqn}
{\bar\nabla}^2 h+\frac{s}{f}{\bar g}^{\mu\nu}\d_\mu f \d_\nu h-\frac{l(l+s-1)}{f^2}h=0.
\eea
As observed in \cite{KodamaOne}, this equation is equivalent to the wave equation on the background (\ref{TensPertNew}) assuming that the scalar function $\Psi$ has the form
\bea
\Psi=h(x)Y_l(y),\quad {\hat\nabla}^2 Y_l=-l(l+s-1)Y_l\,.
\eea
Therefore, to understand the mode $H$, one needs to study the wave equation on the {\it reduced} MP background. As usual, the details are slightly different in even and odd dimensions. 

\bigskip
\noindent
{\bf Even dimensions}

The wave equation for non--degenerate MP geometry is reviewed in section \ref{SecMPwave}, and for separable function $\Psi$ given by (\ref{WaveEqnTwo}), this equation reduces to a system of ODEs (\ref{SeparWaveEvenKerr}). To describe the solutions of equation (\ref{HtensEqn}), we need to set several rotation parameters in (\ref{SeparWaveEvenKerr}) to zero and perform a reduction along the sphere. As discussed in section \ref{SecSubProca}, this always leads to an even--dimensional base, so equation (\ref{WaveEqn}) is replaced by 
\bea\label{AnstzScalSphere}
\Psi=E\Phi(r)\left[\prod_{j}^p X_j(x_j)\right]Y_l(y),\quad E=\exp\left[i\omega t+i\sum_j^p n_j\phi_j\right]\,,
\eea
The metric of the degenerate MP geometry is given by a modified version of (\ref{MPevenInFrames})
\bea
ds^2&=&\left[\frac{1}{FR}{\tilde m}^{(0)}_{+}{\tilde m}^{(0)}_{-}+
\sum_k^p\frac{1}{d_k}{\tilde m}^{(k)}_{+}{\tilde m}^{(k)}_{-}\right]+
\sum_{k=p+1}^n\frac{1}{d_k}{\tilde m}^{(k)}_{+}{\tilde m}^{(k)}_{-}\nn
&=&\left[\frac{1}{FR}{\tilde m}^{(0)}_{+}{\tilde m}^{(0)}_{-}+
\sum_k^p\frac{1}{d_k}{\tilde m}^{(k)}_{+}{\tilde m}^{(k)}_{-}\right]+
\left[r\prod_k^p x_k\right]^2d\Omega_s^2\,,
\eea
where the terms in the square brackets describe the base in (\ref{TensPertNew}). 
Equations (\ref{SeparWaveEvenKerr}) are replaced by
\bea\label{SeparWaveEvenRedS}
&&q(r)\frac{\d}{\d r}\left[\frac{\Delta}{q(r)} \frac{\d\Psi}{\d r}\right]+\frac{R^2}{\Delta}
\left[\omega-
\sum_k\frac{a_k n_k}{r^2+a_k^2}\right]^2\Psi-(-r^2)^{s'}P_{p-1}[-r^2]\Psi=0\,,\nn
&&q(x_i)\frac{\d}{\d x_i}\left[\frac{H_i}{q(x_i)} \frac{\d \Psi}{\d x_i}\right]-
{H_i}{}\left[\omega-
\sum_k\frac{a_k n_k}{a_k^2-x_i^2}\right]^2\Psi+(x_i^2)^{s'}P_{p-1}[x_i^2]\Psi=0\,,\\
&&\d_t\Psi=i\omega\Psi,\quad \d_{\phi_k}\Psi=in_k\Psi\,\quad P_{p-1}[0]=l(l+s-1).\nonumber
\eea
with $q(x)=1$ and $s'=\frac{s}{2}$. Functions $(R,\Delta,H_i)$ are obtained from the general expressions by setting $a_{p+1}=\dots=a_n=0$. 

\bigskip
\noindent
{\bf Odd dimensions}

In odd dimensions, the ansatz (\ref{AnstzScalSphere}) again leads to equations (\ref{SeparWaveEvenRedS}), but this time $q(x)=x$ and $s'=\frac{s+1}{2}$. Equations (\ref{SeparWaveEvenRedS})  replace the system (\ref{SeparWaveOddKerr}) encountered in the non--degenerate case.

\bigskip
\noindent
This concludes our discussion of tensor modes (\ref{TensPertNew}) and their separability on the MP background.

\section{Summary}
\label{SecSumry}

Let us summarize the main results of this article. We considered the MP and GLPP geometries with at least one rotation parameter set to zero. Then the geometry splits into an $s$--dimensional sphere $S^s$ with a nontrivial warp factor and a $(d-s)$--dimensional base. As reviewed in section \ref{SecModes}, in such situations gravitational modes are naturally divided into scalar, vector, and tensor modes on the base, and we focused on the first two options. Here is the summary of our main results:
\begin{itemize}
\item 
Vector modes of gravitational waves coming from a reduction on $S^1$ are discussed in sections \ref{SecVecOdd} and \ref{SecVecEven}. The separable ansatze for such modes are given by (\ref{AnstzMaxw}) or (\ref{AnstzMaxwEvenOne}). The solutions are controlled by one scalar function $\Psi$, defined by (\ref{PsiSeparDef}), which satisfies differential equations 
(\ref{ODEoddMgen})/(\ref{EvenDimGenRed}) in odd/even dimensions. Factors $S$ in these equations are given by (\ref{OddNuParam}) and (\ref{EvenNuParam}). 
\item Vector modes of gravitational waves coming from a reduction on $S^s$ with $s>1$ are discussed in section \ref{SecSubProca}. The separable ansatz is given by equations (\ref{SphereVec}), (\ref{AnstzMaxwProca}),  (\ref{PsiSeparDefProca}). The vector field satisfies a generalized Proca equation (\ref{SnReducMN}), which translates into ODEs (\ref{EvenDimProca})/(\ref{OddDimProca}) in even/odd dimensions. We found that the generalized Proca equation is separable only in two cases shown as options a) and b) on page \pageref{LabBrnchAB}, and they give rise to two options (\ref{ProcaGfact}) for the ``mass terms'' in the ODEs for the scalar $\Psi$. The standard Proca equation can be viewed as a special case of option b), while the gravitational waves are described by a special case (\ref{ProcaZetaNu}) of option a). If all rotation parameters are turned off, then the separable ansatz (\ref{AnstzMaxwProca}) for vector fields no longer works. However, in this case, the MP metric reduces to the Schwarzschild geometry, and the solutions of the relevant Proca equation are briefly discussed in section \ref{SecSubStatic}. 
\item Scalar mode coming from a reduction on a circle is discussed in section \ref{SecSubScalCrcl}, where it is observed that such mode can be removed by a diffeomorphism unless the circle has a trivial warp factor. 
In the context of the rotating geometries, the most interesting situation of nontrivial scalar modes arises in perturbing BH$\times T^p$ geometry, and the relevant solution is given by equation (\ref{PertScalStartV4}).
\item Scalar mode coming from a reduction on a sphere $S^s$ is discussed in section \ref{SecSubScalSph}. Such mode is described by a traceless symmetric tensor under $SO(s+1)$ rotations, so in the literature is it often called the tensor mode when classification is done from the perspective of the sphere. Dynamics of this mode (\ref{TensPertNew}) is governed by differential equations (\ref{SeparWaveEvenRedS}), and the spherical harmonics $T_{ab}(y)$ are reviewed in the Appendix \ref{AppSphHarm}. 
\end{itemize}
\noindent
The four items above cover the vector and scalar modes on the base of partially rotating black holes, and the tensor modes are beyond the scope of this article. Such modes are closely related to gravitational waves on non--degenerate MP and GLPP backgrounds since in this case no reduction is performed, and the entire geometry can be viewed as the ``base''. Some progress towards separating equations for gravitational waves in the vicinity of non--degenerate black hole is reported in section \ref{SecHair}, but unfortunately the full solution is still missing. 

\section{Discussion}

In this article we have demonstrated full separability of dynamical equations governing several polarizations of gravitational waves in the backgrounds of certain MP and GLPP black holes. Considering the situation when at least one rotation parameter vanishes and focusing on vector and scalar modes, we reduced the problem to solving generalized Maxwell, Proca, or wave equations which are obtained from their standard counterparts by inserting additional coordinate--dependent factors. We demonstrated full separability of such equations and analyzed some properties of the resulting ODEs. We also presented some special solutions describing gravitational waves on MP and GLPP geometries with a full set of non--vanishing rotation parameters, but the most general modes in this case are still missing. In contrast to the Teukolsky's procedure for separating equations for gravitons in four dimensions, our construction is  based on projections of metric components rather than curvature tensors. This difference has already been encountered in the past, when Maxwell and Proca equations were separated in 
the MP and GLPP backgrounds using projections of vector fields rather than their field strenghts.

This work can be extended in several directions. First and foremost, it would be very interesting to demonstrate separability of the most general gravitational waves on the MP and GLPP geometries with a full set of non--vanishing rotation parameters. One can also study such waves on the backgrounds which go beyond the minimal gravity, for example, on rotating geometries produced by charged black holes and branes. In this case, linear perturbations of the metric mix with excitations of various fluxes, and separable combinations are expected to be given by some linear superpositions of such modes. Finally, one can use analytical solutions for gravitational waves obtained in this article to study physical properties of gravitational waves in higher dimensions, such as the detailed structure of their quasinormal nodes.

\section*{Acknowledgements}

This work was supported in part by the DOE grant DE-SC0015535.

\appendix

\section{Rotating black holes and their excitations}
\label{AppMP}

In this appendix we summarize some well-known properties of the Myers--Perry geometry \cite{MyersPerry} and its extension to the GLPP black hole \cite{GLPP} with a negative cosmological 
constant. In section \ref{SecSubMP} we present the metrics, the Killing--Yano tensors, and special frames leading to separation of variables in various dynamical equations. Such separation for scalars, vectors, and antisymmetric tensors is reviewed in section \ref{SecMPwave}, and separability of gravitational waves is proved in the in the main body of the article using the properties of the geometries reviewed in this appendix.

\subsection{Myers--Perry black hole and its symmetries}
\label{SecSubMP}

In this subsection we will review some well-known properties of the $d$--dimensional Myers--Perry and GLPP geometries \cite{MyersPerry,GLPP}. Given a complicated form of the GLPP metric supported by a negative cosmological constant, it is instructive to start with a relatively simple Myers--Perry case and introduce the modifications caused by the cosmological constant only in the final formulas. 

\bigskip

\noindent
{\bf Even dimensions}

The form of metric of rotating black holes differs between even and odd values of $d$, so we begin with quoting the Myers--Perry solution in $d=2n+2$ dimensions \cite{MyersPerry,Myers}:
\bea\label{MPeven}
ds^2&=&-dt^2+\frac{Mr}{FR}\Big(dt+\sum_{i=1}^n a_i\mu_i^2 d\phi_i\Big)^2+\frac{FR dr^2}{R-Mr}
+\sum_{i=1}^n(r^2+a_i^2)\Big(d\mu_i^2+\mu_i^2 d\phi_i^2\Big)\nn
&&+r^2d\alpha^2.
\eea
Here variables $(\mu_i,\alpha)$ are subject to a constraint
\bea
\alpha^2+\sum_{i=1}^n\mu_i^2=1,
\eea
and functions $F$, $R$ are defined by 
\bea\label{MPdefFR}
F=1-\sum_{k=1}^n\frac{a_k^2\mu_k^2}{r^2+a_k^2},\quad R=\prod_{k=1}^n (r^2+a_k^2).
\eea
The geometry (\ref{MPeven}) is highly symmetric: it admits $(n+1)$ Killing vectors corresponding to translations in $(t,\phi_i)$ directions, as well as a family of Killing--Yano 
tensors (KYT) \cite{Kub1w1,Kub1w2,Kub1w3,Kub1w4,Kub1w5,Kub1w6,Kub2w1,Kub2w2,
Kub2w3,Kub3w1,
Kub3w2,Kub3w3,ChLKill,KubRev}. 
To describe the latter, it is convenient to replace $(r,\mu_1\dots \mu_n)$ by new coordinates $(x_0,x_1\dots x_n)$ defined by
\bea\label{Xmap}
x_0=r^2,\quad (a_i\mu_i)^2=\frac{1}{c_i^2}\prod_1^n(a_i^2-x_k^2),\quad c_i^2=\prod(a_i^2-a_k^2).
\eea
Without loss of generality, the rotation parameters and the ellipsoidal coordinates $x_j$ can be ordered as 
\bea\label{XAorderEven}
x_n\le a_n\le \dots x_1 \le a_1\,.
\eea
The general equation for an anti--symmetric KYT \cite{YanoW1,YanoW2,YanoW3} is 
\bea
\nabla_\mu Y_{\nu_1\dots \nu_p}+\nabla_{\nu_1} Y_{\mu\nu_2\dots \nu_p}=0,
\eea
and for the Myers--Perry black holes (\ref{MPeven}), all Killing--Yano tensors are given by a very elegant formula \cite{Kub1w1,Kub1w2,Kub1w3,Kub1w4,Kub1w5,Kub1w6,Kub2w1,Kub2w2,Kub2w3,Kub3w1,
Kub3w2,Kub3w3}
\bea\label{Kub2}
Y^{2(n-k)}=\star\left[\wedge h^k\right].
\eea
Here $h$ is a special two--form that will be written below. As demonstrated in \cite{LMaxw,LBfield}, the eigenvectors of the antisymmetric tensor $h$ play the central role in separation of Maxwell's equations and their generalizations to higher anti--symmetric forms, and in this article we extend such separation to gravitational waves. Using the eigenvectors of a symmetric tensor $t_{\mu\nu}=h_{\mu\alpha}{h_\nu}^\alpha$ as frames, one finds very simple expressions for the two--form $h$ and for the metric \cite{Kub1w1,Kub1w2,Kub1w3,Kub1w4,Kub1w5,Kub1w6,Kub2w1,Kub2w2,Kub2w3,Kub3w1,
Kub3w2,Kub3w3,UniqueKYw1,UniqueKYw2,
UniqueKYw3,UniqueKYw4}:
\bea\label{HInFrame}
h=re^r\wedge e^t+\sum_i x_i e^{x_i}\wedge e^i,\quad 
ds^2=-(e^t)^2+(e^r)^2+\sum_k [(e^{x_k})^2+(e^k)^2].
\eea
The frames were constructed in \cite{Kub1w1,Kub1w2,Kub1w3,Kub1w4,Kub1w5,Kub1w6,Kub2w1,Kub2w2,Kub2w3,Kub3w1,
Kub3w2,Kub3w3}, and we will write only the components with upper indices, $e_A^\mu$, using the notation introduced in \cite{ChLKill}:
\bea\label{AllFramesMP}
e_t&=&-\sqrt{\frac{R^2}{FR\Delta}}\left[\d_t-
\sum_k\frac{a_k}{r^2+a_k^2}\d_{\phi_k}\right],\quad 
e_r=\sqrt{\frac{\Delta}{FR}}\d_r,\nn
e_i&=&-\sqrt{\frac{H_i}{d_i}}\left[\d_t-\sum_k\frac{a_k}{a_k^2-x_i^2}\d_{\phi_k}
\right],\quad e_{x_i}=\sqrt{\frac{H_i}{d_i}}\d_{x_i}\,.
\eea
Here we introduced convenient expressions involving $x$ coordinates defined by (\ref{Xmap}),
\bea\label{MiscElliptic}
 H_i=\prod_k(a_k^2-x_i^2),\quad d_i=(r^2+x_i^2)\prod_{k\ne i}(x_k^2-x_i^2)\,,
\eea
as well as their radial counterparts
\bea\label{MisclEllipticEvev}
R=\prod_{k} (r^2+a_k^2),\quad
FR=\prod_k(r^2+x_k^2),\quad \Delta=R-Mr.
\eea
Functions $(R,FR)$ can be interpreted as $H_0$ and $d_0$ with $x_0=i r$. 

Note that, apart from the common overall factors, the components $(e_t^\mu,e_r^\mu)$ of the frames depend only on $r$, while the components $(e_i^\mu,e_{x_i}^\mu)$ depend only on $x_i$. This crucial fact is responsible for separation of variables in all equations discussed in this article. Interestingly, frames (\ref{AllFramesMP}) admit a very simple extension covering the GLPP geometry, even though the metric itself is much more complicated than the one for the Myers--Perry black hole. Specifically, to add the cosmological constant to frames (\ref{AllFramesMP}), one should make the following replacements \cite{Kub1w1,Kub1w2,Kub1w3,Kub1w4,Kub1w5,Kub1w6,Kub2w1,Kub2w2,Kub2w3,Kub3w1,
Kub3w2,Kub3w3}
\bea\label{AllFramesGLPP}
e_t\rightarrow \frac{1}{Q_r}e_t,\quad 
e_r\rightarrow Q_r e_r,\quad
e_j\rightarrow \frac{1}{Q_j}e_j,\quad e_{x_j}\rightarrow Q_j e_{x_j}\,.
\eea
Here factors $(Q_r,Q_j)$ depend on the cosmological parameter $L$ introduced in \cite{GLPP} as
\bea\label{QfactDef}
Q_r=\left[1-L r^2\frac{R}{\Delta}\right]^{\frac{1}{2}},\quad Q_j=\sqrt{1+L x_j^2}\,.
\eea

As found in \cite{LMaxw} and reviewed in section \ref{SecMPwave}, separation of the Maxwell's equations occurs not for components $(A_t,A_r,A_{\phi_i},A_{x_i})$, but for projections of the gauge fields to particular combinations of the frames (\ref{AllFramesMP}) or (\ref{AllFramesGLPP}). Specifically, it is convenient to define $m^{(I)}_\pm$ by
\bea\label{Mframes}
&&m^{(0)}_\pm\equiv \sqrt{FR}(e_r\mp e_t)
=\frac{R}{\sqrt{\Delta}}\left\{\frac{Q_r\Delta}{R}\d_r\pm 
\frac{1}{Q_r}\left[\d_t-
\sum_k\frac{a_k}{r^2+a_k^2}\d_{\phi_k}\right]\right\},\nn
&&m_\pm^{(j)}\equiv\sqrt{d_i}(e_{x_i}\mp ie_i)
=\sqrt{{H_j}}\left\{Q_j\d_{x_j}\pm \frac{i}{Q_j}\left[\d_t-\sum_k\frac{a_k}{a_k^2-x_j^2}\d_{\phi_k}
\right]\right\}\,.
\eea
Then, as demonstrated in \cite{LMaxw}, it is the set of projections $[m^{(I)}_\pm]^\mu A_\mu$ that separates. Note that the components $m_\pm^{(j)\mu}$ depend only on $x_j$, and $m_\pm^{(0)\mu}$ depend only on $r$. The metric and the generator $h$ of the Killing--Yano tensors (\ref{Kub2})--(\ref{HInFrame}) can be written as 
\bea\label{MPevenInFrames}
ds^2&=&\frac{1}{FR}{\tilde m}^{(0)}_{+}{\tilde m}^{(0)}_{-}+
\sum_k\frac{1}{d_k}{\tilde m}^{(k)}_{+}{\tilde m}^{(k)}_{-},\quad {\tilde m}^{(I)}_\pm\equiv  [{m}^{(I)}_\pm]_\mu dx^\mu\\
h&=&\frac{r}{2FR}{\tilde m}^{(0)}_+\wedge {\tilde m}^{(0)}_-+\sum_{k=1}^n \frac{x_k}{2id_k}
{\tilde m}^{(k)}_{+}\wedge{\tilde m}^{(k)}_{-}\,.\nonumber
\eea
The frames (\ref{Mframes}) will play an important role throughout this article. 

\bigskip

\noindent
{\bf Odd dimensions}

Let us now briefly discuss the Myers--Perry and GLPP black holes in odd dimensions ($d=2n+3$). The MP metric is given by \cite{MyersPerry,Myers}
\bea\label{MPodd}
\hskip -0.3cm
ds^2=-dt^2+\frac{Mr^2}{FR}\Big(dt+\sum_{i=1}^{n+1} a_i\mu_i^2 d\phi_i\Big)^2+\frac{FR dr^2}{R-Mr^2}
+\sum_{i=1}^{n+1}(r^2+a_i^2)\Big(d\mu_i^2+\mu_i^2 d\phi_i^2\Big),
\eea
and coordinates $\mu_i$ are subject to a constraint
\bea
\sum_{i=1}^{n+1}\mu_i^2=1.\nonumber
\eea
The counterparts of the special frames (\ref{HInFrame}), (\ref{AllFramesGLPP}) are given by \cite{Kub1w1,Kub1w2,Kub1w3,Kub1w4,Kub1w5,Kub1w6,ChLKill}
\bea\label{AllFramesMPOdd}
e_t&=&-\frac{1}{Q_r}\sqrt{\frac{R^2}{FR\Delta}}\left[ \d_t
-\sum_k\frac{a_k}{r^2+a_k^2}\d_{\phi_k}\right],\quad e_r=Q_r\sqrt{\frac{\Delta}{FR}}\d_r,\quad
e_{x_i}=Q_i\sqrt{\frac{H_i}{x_i^2 d_i}}\d_{x_i}\nn
e_i&=&-\frac{1}{Q_i}\sqrt{\frac{H_i}{x^2_id_i}}\left[\d_t-\sum_k\frac{a_k}{a_k^2-x_i^2}\d_{\phi_k}
\right],\quad 
e_\psi=-\frac{\prod a_i}{r\prod x_k}\left[\d_t-\sum_k\frac{1}{a_k}\d_{\phi_k}
\right].
\eea
Relations (\ref{MiscElliptic}) still hold, while their radial counterparts (\ref{MisclEllipticEvev}) are replaced by
\bea\label{DelktaOdd}
R=\prod_{k} (r^2+a_k^2),\quad FR=r^2\prod_k(r^2+x^2_k),\quad \Delta=R-Mr^2.
\eea
The ordering of rotation parameters and ellipsoidal coordinates $x_j$ is slightly different from (\ref{XAorderEven}):
\bea\label{XAorderOdd}
a_{n+1}\le x_n\le \dots \le a_1\,.
\eea
The Killing--Yano tensors still have the form (\ref{Kub2})--(\ref{HInFrame}), although the metric acquires an extra term $(e_\psi)^2$, and we refer to \cite{ChLKill} for the detailed discussion. Finally, the light--cone frames $m^{(I)}_\pm$ are given by the counterparts of (\ref{Mframes}), and there is an additional frame $n^\mu$:
\bea\label{GenFramesOddD}
&&\hskip -1cm\left[m_\pm^{(0)}\right]^\mu\d_\mu=\frac{R}{\sqrt{\Delta}}\left\{\frac{Q_r\Delta}{R}\d_r\pm \frac{1}{Q_r}\left[\d_t-
\sum_k\frac{a_k}{r^2+a_k^2}\d_{\phi_k}\right]\right\},\qquad \Delta=R-Mr^2,\\
&&\hskip -1cm\left[m_\pm^{(j)}\right]^\mu
\d_\mu=\sqrt{{H_j}}\left\{Q_j\d_{x_j}\pm \frac{i}{Q_j}\left[\d_t-\sum_k\frac{a_k}{a_k^2-x_j^2}\d_{\phi_k}
\right]\right\}\,,\quad
n^\mu\d_\mu=\d_t-\sum_k\frac{1}{a_k}\d_{\phi_k}.\nonumber
\eea
Note that the components $[m_\pm^{(j)}]^\mu$ depend only on $x_j$, $[m_\pm^{(0)}]^\mu$ depend only on $r$, and $n^\mu$ are constants. This feature is crucial for ensuring separation of variables in various dynamical equations reviewed in the next subsection. To study such equations, we will need the expression for the inverse metric in terms of these frames: 
\bea\label{GinvOddApp}
g^{\mu\nu}\d_\mu\d_\nu&=&\frac{1}{FR}\left[m_+^{(0)}\right]^\mu 
\left[m_-^{(0)}\right]^\nu\d_\mu\d_\nu+
\sum_{j=1}^n \frac{1}{x_j^2 d_j}{\left[m_+^{(j)}\right]^\mu\left[m_-^{(j)}\right]^\mu\d_\mu\d_\nu}\\
&&+
\left[\frac{\prod a_i}{r\prod x_k}\right]^2n^\mu n^\nu\d_\mu\d_\nu\,.\nonumber
\eea
Expressions (\ref{MPevenInFrames}) and (\ref{GenFramesOddD}) for the frames are used throughout this article.

\subsection{Separation of dynamical equations}
\label{SecMPwave}

In this subsection we review separation of variables in various dynamical equations in the Myers--Perry and GLPP geometries. We focus on the general case of 
${\bar n}=\left[\frac{d-1}{2}\right]$ distinct non--zero rotation parameters\footnote{Recall that we use ${\bar n}$ to denote the maximal number of rotation parameters, and we reserve $n$ for the number of non--cyclic angular coordinates in the MP and GLPP geometries. One has $n={\bar n}$ in even dimensions and ${\bar n}=n+1$ otherwise.}, and some degenerate situations are discussed in the main body of this article in the context of gravitational waves. As in the previous subsection, we present separate discussions of even and odd dimensions. 

\bigskip

\noindent
{\bf Even dimensions}

We begin with analyzing the wave equation
\bea\label{WaveEqn}
\frac{1}{\sqrt{-g}}\d_\mu\left[\sqrt{-g}g^{\mu\nu}\d_\nu \Psi\right]=0
\eea
in the background (\ref{MPeven}) and its GLPP extension. Imposing a separable ansatz
\bea\label{WaveEqnTwo}
\Psi=E\Phi(r)\left[\prod X_i(x_i)\right],\qquad E=e^{i\omega t+i\sum n_i\phi_i}\,,
\eea
and substituting in into (\ref{WaveEqn}), we conclude that the function $\Psi$ satisfies a system of ordinary differential equations \cite{LMaxw}:
\bea\label{SeparWaveEvenKerr}
&&\frac{\d}{\d r}\left[\Delta Q_r^2\frac{\d\Psi}{\d r}\right]+\frac{R^2}{Q_r^2\Delta}
\left[\omega-
\sum_k\frac{a_k n_k}{r^2+a_k^2}\right]^2\Psi-P_{n-1}[-r^2]\Psi=0\,,\\
&&\frac{\d}{\d x_i}\left[H_i Q_i^2\frac{\d \Psi}{\d x_i}\right]-
\frac{H_i}{Q_i^2}\left[\omega-
\sum_k\frac{a_k n_k}{a_k^2-x_i^2}\right]^2\Psi+P_{n-1}[x_i^2]\Psi=0\,,\nn
&&\d_t\Psi=i\omega\Psi,\quad \d_{\phi_k}\Psi=in_k\Psi\,.\nonumber
\eea
Here $P_{n-1}[y]$ is an arbitrary polynomial of degree $(n-1)$, and it is crucial that all equations (\ref{SeparWaveEvenKerr}) contain {\it the same} function $P_{n-1}$. Therefore, the separable ansatz (\ref{WaveEqnTwo}) reduces one PDE (\ref{WaveEqn}) to a system of $d=2(n+1)$ ordinary differential equations for function $\Psi$, and as expected, the resulting solution depends on $d-1=2n+1$ integration constants:
\label{CountWave}
\begin{itemize}
\item $(n+1)$ parameters $(\omega,n_i)$;
\item $n$ free coefficients of the polynomial $P_{n-1}$.
\end{itemize}
This guarantees full integrability of the wave equation.

Let us now review separability of Maxwell's equations in the Myers--Perry and GLPP backgrounds. 
As demonstrated in \cite{LMaxw}, the most general separable ansatz for the Maxwell field in even dimensions has the form\footnote{See also \cite{KubMaxwW1,KubMaxwW2,KubMaxw1w1,KubMaxw1w2,KubMaxw1w3} for further explorations of this ansatz.}
\bea\label{MaxwAnstz}
[m^{(I)}_\pm]^\nu A_\nu=\left[\zeta+\frac{x_I}{x_I\pm\mu}\right][m^{(I)}_\pm]^\nu\d_\nu \Psi,\quad x_0=-ir,
\eea
where the scalar function $\Psi$ is given by (\ref{WaveEqnTwo}). Relations (\ref{MaxwAnstz}) have a free parameter $\mu$ that will play a role of one of the separation constants. Equations (\ref{MaxwAnstz}) can be easily solved for the gauge potential,
\bea\label{MaxwAnsAlt}
A_\mu={K_\mu}^\nu\d_\nu\Psi,\quad 
{K_\mu}^\nu=-\sum_I\sum_{\alpha=\pm}\frac{i\alpha}{x_I+\alpha\mu}\frac{1}{d_I}
[m^{(I)}_{-\alpha}]_\mu
[m^{(I)}_\alpha]^\nu\,,
\eea
but we find the expression (\ref{MaxwAnstz}) to be more compact and useful for performing calculations. Substitution of the ansatz (\ref{MaxwAnstz}) into Maxwell's equations leads to a system of ODEs \cite{LMaxw}:
\bea\label{MaxwEvenDim}
&&{D_j}\frac{\d}{\d x_j}\left[\frac{Q_j^2 H_j}{D_j}\d_{x_j}\Psi\right]+\left\{\frac{2\Lambda}{D_j}-
\frac{H_j W_j^2}{Q_j^2}-
\Lambda +P_{n-2}[x_j^2] D_j\right\}\Psi=0,
\nn
&&{D_r}\frac{\d}{\d r}\left[\frac{Q_r^2\Delta}{D_r}\d_r\Psi\right]-\left\{\frac{2\Lambda}{D_r}-
\frac{R^2 W_r^2}{Q_r^2\Delta}-
\Lambda +P_{n-2}[-r^2]  D_r\right\}\Psi=0\,,\\
&&\d_t\Psi=i\omega\Psi,\quad \d_{\phi_k}\Psi=in_k\Psi\,.\nonumber
\eea
Various functions and constants appearing in these equations are defined by
\bea\label{WfuncMaxw}
&&\Omega=\omega-\sum\frac{n_ka_k}{\Lambda_k},\quad 
W_j=\omega-\sum\frac{n_k a_k}{a_k^2-x_j^2},\quad W_r=\omega-\sum\frac{n_k a_k}{a_k^2+r^2}\,,\\
&&
D_j=1-\frac{x_j^2}{\mu^2}\,,\quad D_r=1+\frac{r^2}{\mu^2}\,,\quad
\Lambda=\frac{\Omega}{\mu}
\prod \Lambda_k,\quad \Lambda_i=(a_i^2-{\mu^2})\,.
\nonumber
\eea
As in the scalar case, the solutions are specified by $(2n+1)$ integration constants, which can be identified as
\begin{itemize}
\item $(n+1)$ parameters $(\omega,n_i)$;
\item $(n-1)$ free coefficients of the polynomial $P_{n-2}$;
\item constant $\mu$.
\end{itemize}
As demonstrated in \cite{LMaxw}, the ansatz (\ref{MaxwAnstz}) covers all $d-2$ polarizations of the electromagnetic field: they correspond to different ranges of $\mu$. Note that the ansatz (\ref{MaxwAnstz}) contains some residual gauge freedom controlled by a free parameter $\zeta$, and in \cite{KubMaxwW1,KubMaxwW2} it was observed that by setting $\zeta=-1$, one recovers the Lorenz gauge,
\bea\label{LornzGauge}
\nabla_\mu A^\mu=0.
\eea
Remarkably, the ansatz (\ref{MaxwAnstz}) with $\zeta=-1$ ensures separability of the Proca equation as well \cite{Proca}, and equations (\ref{MaxwEvenDim}) contain some additional terms containing the mass of the gauge field. We discuss the Proca equation and its generalizations in section \ref{SecSubProca}. 

\bigskip

Going beyond scalar fields and one--form potentials, one can look at a generalized Maxwell's equations for $p$--form fields,
\bea\label{FieldEqn}
d\star dA^{(p)}=0.
\eea 
Remarkably, symmetries of the MP and GLPP geometries are sufficiently rich to ensure separability of such equations as well \cite{LBfield}. As an example, we quote the separable ansatz for the two--form potential found in \cite{LBfield}:
\bea\label{BfieldAnswSpec}
&&\sum_{\mu,\nu}[m^{(I)}_\alpha]^\mu [m^{(J)}_\beta]^\nu A^{(2)}_{\mu\nu}=
(x^2_I-x^2_J)\frac{V_{I,\alpha}}{P_I}\frac{V_{J,\beta}}{P_J}
\sum_{i,j}[m^{(I)}_\alpha]^i[m^{(J)}_\beta]^j \d_i\d_j\Psi\,,\\
&&P_I=a x_I^4+bx_I^2+c,\quad V_{I,\alpha}=px_I^2+q+\alpha x_I.\nonumber
\eea
\bea\label{6DanswCrct}
m^{I\mu}_\alpha m^{J\nu}_\beta A_{\mu\nu}=
(x_I^2-x_J^2)\, h^I_\alpha h^J_\beta\, m^{I\mu}_\alpha m^{J\nu}_\beta \d_\mu\d_\nu\Psi,
\quad
h^I_\pm=\frac{1}{e_1+e_2 x_I^2\pm ie_3 x_I}\,.
\eea
Here $\Psi$ is a separable function that has the structure (\ref{WaveEqnTwo}) that satisfies equations similar to (\ref{MaxwEvenDim}). We refer \cite{LBfield} for the explicit form of these equations and for extensions of the separable ansatz to gauge potentials $A^{(p)}$ with $p>2$.

\bigskip
{\bf Odd dimensions}

Let us now discuss dynamical equations in odd dimensions. Starting with a separable function 
$\Psi$ (\ref{WaveEqnTwo}) and substituting it into the wave equation (\ref{WaveEqn}), one finds the odd--dimensional counterpart of ODEs (\ref{SeparWaveEvenKerr}):
\bea\label{SeparWaveOddKerr}
&&r\frac{d}{dr}\left[\frac{Q_r^2\Delta}{r}\frac{d\Phi}{dr}\right]+\frac{R^2}{Q_r^2\Delta}
\left[\omega-
\sum_k\frac{a_k n_k}{r^2+a_k^2}\right]^2\Phi-P_{n-1}[-r^2]\Phi=0,\nn
&&{x_i}\frac{d}{dx_i}\left[\frac{Q_i^2H_i}{x_i}\frac{dX_i}{dx_i}\right]-
\frac{H_i}{Q_i^2}\left[\omega-
\sum_k\frac{a_k n_k}{a_k^2-x_i^2}\right]^2X_i+P_{n-1}[-x_i^2]X_i=0\,, \\
&&\d_t\Psi=i\omega\Psi,\quad \d_{\phi_k}\Psi=in_k\Psi\,.\nonumber
\eea
As before, the relations (\ref{WaveEqnTwo}), (\ref{SeparWaveOddKerr}) guarantee full separability of the wave equation, and solution depends on $d-1=2n+2$ free parameters:
\begin{itemize}
\item $(n+2)$ parameters $(\omega,n_i)$;
\item $n$ free coefficients of the polynomial $P_{n-1}$.
\end{itemize}

The counterpart of the ansatz (\ref{MaxwAnstz}) for odd dimensions was also found in \cite{LMaxw}, where it was shown that the most general separable solution of the Maxwell's equations has the form
\bea\label{NewAnstzOdd}
[m^{(I)}_\pm]^\nu A_\nu=\mp\frac{i}{x_I\pm\mu}[m^{(I)}_\pm]^\nu\d_\nu \Psi,\quad
n^\nu A_\nu=-\frac{i}{\mu}n^\nu\d_\nu\Psi\,,
\eea
with $\Psi$ given by (\ref{WaveEqnTwo}). The resulting ordinary differential equations are \cite{LMaxw}
\bea\label{OddMPmaster}
&&\frac{D_j}{x_j}\frac{d}{dx_j}\left[\frac{Q_j^2H_j}{x_j D_j}X_j'\right]+\left\{\frac{2\Lambda}{D_j}-
\frac{H_jW_j^2}{x_j^2Q_j^2}+\frac{\AA D_j}{x_j^2}{\tilde\Omega}^2+P_{n-2}[-x_j^2]D_j\right\}=0,\nn
&&\frac{D_r}{r}\frac{d}{dr}\left[\frac{Q_r^2\Delta}{r D_r}{\dot\Phi}\right]+\left\{\frac{2\Lambda}{D_r}+
\frac{R^2W_r^2}{r^2Q_r^2\Delta}-\frac{\AA D_r}{r^2}{\tilde\Omega}^2+P_{n-2}[r^2] D_r\right\}\Phi=0.
\eea
Here functions $(D_j,D_r,W_j,W_r)$ are still given by (\ref{WfuncMaxw}), and the constants 
$(\AA,{\tilde\Omega},\Lambda)$ are defined as
\bea\label{AAdefApp}
\AA=\left[\prod a_k\right]^2,\quad
{\tilde\Omega}=\omega-\sum_k \frac{n_k}{a_k}\,,\quad
\Lambda=\frac{\Omega}{\mu^3}
\prod \Lambda_k\,.
\eea
As in even dimensions, one can show that the system (\ref{NewAnstzOdd})--(\ref{OddMPmaster}) describes the most general separable solution of the Maxwell's equations, and it depends on $d-1$ separation parameters and covers $d-2$ independent polarizations \cite{LMaxw}. 

\bigskip

Separation of equations  (\ref{FieldEqn}) for higher forms follows the same logic as in even dimensions. For example, the counterparts of equations (\ref{BfieldAnswSpec}) for the two--form potential are
\bea\label{Odd2Anstz}
m^{I\mu}_\alpha m^{J\nu}_\beta A_{\mu\nu}&=&
(x_I^2-x_J^2)\, h^I_\alpha h^J_\beta\, m^{I\mu}_\alpha m^{J\nu}_\beta \d_\mu\d_\nu\Psi,\\
m^{I\mu}_\alpha n^{\nu} A_{\mu\nu}&=&
x_I^2\, h^I_\alpha h^J_\beta\, m^{I\mu}_\alpha n^{\nu}\, \d_\mu\d_\nu\Psi\,.\nonumber
\eea
We refer to \cite{LBfield} for further details and for the discussion of higher forms.

\section{Spherical harmonics}
\label{AppSphHarm}

In this appendix we summarize some properties of spherical harmonics which are used in the main body of the article. As reviewed in section \ref{SecModes}, reduction of gravitational waves on spheres gives rise to scalars, vectors, and symmetric rank--two tensors, so we will focus only on these types of spherical harmonics. 

\bigskip
\noindent
{\bf Scalar harmonics}

We begin with reviewing the scalar spherical harmonics satisfying the Helmholtz equation on $S^s$:
\bea\label{ScalHarmOne}
{\hat\nabla}^2 Y+\la_S Y=0.
\eea
Such harmonics can be constructed by studying solutions of the Laplace equations in an auxiliary $(s+1)$--dimensional flat space with a metric
\bea\label{FlatForHarm}
ds^2=dr^2+r^2 d\Omega_s^2=dx_mdx_m\,.
\eea
To construct the harmonic functions of the form $\Psi=r^k Y(y)$, we consider homogeneous polynomials of degree $k$:
\bea\label{SphScalPolyn}
\Psi_k=S_{m_1\dots m_k}x^{m_1}\dots x^{m_k}=r^k S_{m_1\dots m_k}y^{m_1}\dots y^{m_k}
\eea
In spherical coordinates, the Laplace equation reads
\bea
\nabla^2 \Psi_k&=&\frac{1}{r^s}\d_r(r^s \d_r\Psi_k)+\frac{1}{r^2}{\hat\nabla}^2 \Psi_k=
\frac{1}{r^2}\left[k(k+s-1)\Psi_k+{\hat\nabla}^2 \Psi_k\right]=0,
\eea
and therefore it reduces to (\ref{ScalHarmOne}) with 
\bea\label{ScalHarmOneEigen}
\lambda_S=k(k+s-1).
\eea 
Written in Cartesian coordinates, the Laplace equation implies tracelessness of the symmetric tensor $S_{m_1\dots m_k}$:
\bea\label{SphHarmScalWave}
{S^m}_{mm_3\dots m_k}=0.
\eea
This condition can be used to count modes with eigenvalues (\ref{ScalHarmOneEigen}):
\bea
{N}_k=\frac{(s+k)!}{k! s!}-\frac{(s+k-2)!}{(k-2)!s!}\,.
\eea
For the two--dimensional sphere, one recovers the well-known $(2k+1)$ spherical harmonics with eigenvalue $\lambda_S=k(k+1)$. 

\bigskip
\noindent
{\bf Vector harmonics}

Let us now consider vector fields on a sphere satisfying the Helmholtz--type equation
\bea\label{VecHarmOne}
{\hat\nabla}^2 V_a+\la_V V_a=0.
\eea
Some solutions of this type can be obtained by applying the derivative $\d_a$ to a scalar harmonic, and to eliminate this degenerate case, we define
\bea
{\tilde V}_a=V_a +{\tilde\la}{\hat\nabla}_a({\hat\nabla}_b V^b)
\eea
The second term describes the redundancy associated with a degenerate solution. Computing divergence of the last expression, one finds
\bea
{\hat\nabla}_a{\tilde V}^a=\left[1+{\tilde\la}{\hat\nabla}^2\right]({\hat\nabla}_b V^b)=
\left[1-{\tilde\la}\la_S\right]({\hat\nabla}_b V^b),
\eea
where $\la_S$ is the eigenvalue of the Laplacian corresponding to the scalar $({\hat\nabla}_b V^b)$. Therefore, one can eliminate the redundancy in ${\tilde V}_a$ by choosing the unique ${\tilde\la}$ that gives
\bea
{\hat\nabla}_a{\tilde V}^a=0
\eea
Making this choice and dropping the tilde, we arrive at the system of equations for non--degenerate vector modes:
\bea\label{SphVecEqnSum}
{\hat\nabla}^2 V_a+\la_V V_a=0,\quad {\hat\nabla}_a{V}^a=0.
\eea
To solve these equations, we look at vectors in the $(s+1)$--dimensional flat space (\ref{FlatForHarm}). The vector counterpart of the expansion (\ref{SphScalPolyn}) reads
\bea
V_m=\sum V_{m;m_1\dots m_k}x^{m_1}\dots x^{m_k}\,,
\eea
and to eliminate the radial component of the vector, we impose a constraint
\bea
x^m V_m=0.
\eea
Using the relation $\d_r V_m=\frac{k}{r}V_m$, we find the Laplacian of the vector in spherical coordinates:
\bea\label{SphHarmCalc}
\nabla^2 V_m&=&\nabla^\mu[\d_\mu V_m-\Gamma_{\mu m}^k V_k]\nn
&=&
g^{\mu\nu}\d_\nu[\d_\mu V_m-\Gamma_{\mu m}^k V_k]-
g^{\mu\nu}\Gamma_{\nu\mu}^\sigma [\d_\sigma V_m-\Gamma_{\sigma m}^k V_k]
-g^{\mu\nu}\Gamma_{\nu m}^\sigma [\d_\mu V_\sigma-\Gamma_{\mu \sigma}^k V_k]\nn
&=&\frac{1}{r^2}{\hat\nabla}^2 V_m+[\d_r-
g^{\mu\nu}\Gamma_{\nu\mu}^r][\d_r V_m-\Gamma_{r m}^k V_k]
-\Gamma_{r m}^\sigma [\d_r V_\sigma-\Gamma_{r\sigma}^k V_k]
-g^{pl}\Gamma_{l m}^r [-\Gamma_{p r}^k V_k]\nn
&=&\frac{1}{r^2}{\hat\nabla}^2 V_m+[\d_r+\frac{s}{r}][\d_r V_m-\frac{1}{r} V_m]
-\frac{1}{r} [\d_r V_m-\frac{1}{r} V_m]
-\frac{1}{r^2}V_m\\
&=&\frac{1}{r^2}{\hat\nabla}^2 V_m+\frac{1}{r^2}V_m[(k-1+s-1)(k-1)-1].\nonumber
\eea
Therefore, the Laplace equation in the $(s+1)$--dimensional flat space,
\bea\label{EqnHarmVec}
\nabla^2 V_m=0,
\eea
implies the first relation in (\ref{SphVecEqnSum}) with 
\bea\label{SphVecEigen}
\la_V=(k-1+s-1)(k-1)-1.
\eea
Furthermore, the condition $V_r=0$ and the second relation in (\ref{SphVecEqnSum}) imply that 
\bea
\nabla_m V^m=\frac{1}{\sqrt{g}}\d_r(\sqrt{g}V^r)+\frac{1}{r^2}{\hat\nabla}_\alpha V^\alpha=0.
\eea
Therefore, in Cartesian coordinates, the Laplace equation and other constraints on vector $V_m$ lead to a set of {\it algebraic equations for constant factors}:
\bea\label{SphVecConstr}
{V_{m;m_1\dots m_{k-2}n}}^{n}=0,\quad {V_{n;m_1\dots m_{k-1}}}^{n}=0,\quad
{V_{m;m_1m_2\dots m_{k}}}x^m x^{m_1}\dots x^{m_k}=0.
\eea
To satisfy the first relation, we need to have $k\ge 2$, so the expression (\ref{SphVecEigen}) for the eigenvalues is often written as 
\bea\label{SphVecEigenOne}
\la_V=\ell (\ell+s-1)-1,\quad \ell=1,2,\dots
\eea
For $\ell=1$, the constraints (\ref{SphVecConstr}) imply that $V_m$ is a Killing vector. 

\bigskip
\noindent
{\bf Tensor harmonics}

The tensor harmonics satisfying the Helmholtz--type equation
\bea\label{TensHarmOne}
{\hat\nabla}^2 T_{ab}+\la_T T_{ab}=0
\eea
can contain the contributions from scalars (${\hat\nabla}_a {\hat\nabla}_b S$) and from vectors 
 (${\hat\nabla}_{(a} V_{b)}$), and using the logic that led to (\ref{SphVecEqnSum}), one can eliminate  these terms by imposing constraints
\bea\label{TensHarmTwo}
{\hat\nabla}_a T^{ab}=0,\quad {T_a}^a=0.
\eea
As in the scalar and vector case, we define a tensor in the $(s+1)$--dimensional flat space (\ref{FlatForHarm}) by introducing
the expansion
\bea
T_{mn}=T_{mn;p_1\dots p_k}x^{p_1}\dots x^{p_k}\,,
\eea
and we eliminate the radial component by imposing the constraint
\bea
T_{mn}x^m=0. 
\eea
Then a straightforward extension of the derivative (\ref{SphHarmCalc}) gives
\bea\label{SphHarmCalcTens}
\nabla^2 T_{mn}=\frac{1}{r^2}{\hat\nabla}^2 T_{mn}+\frac{1}{r^2}T_{mn}[(k-2+s-1)(k-2)-2].
\eea
Imposing the Laplace equation in the flat $(s+1)$--dimensional space,
\bea
\nabla^2 T_{mn}=0,
\eea
we find the eigenvalues of the tensor Laplacian on the sphere:
\bea\label{SphTensEigen}
\la_T=(k-2+s-1)(k-2)-2.
\eea
The eigenfunctions must satisfy the counterparts of equations (\ref{SphVecConstr}) obtained by rewriting the Laplace equation and various constraints in terms of Cartesian components:
\bea\label{SphTensConstr}
&&{T_{mn;m_1\dots m_{k-2}p}}^{p}=0,\quad {{T^n}_{n;m_1\dots m_k}}=0,\quad 
{T_{np;m_1\dots m_{k-1}}}^{p}=0,\nn
&&{T_{mn;m_1m_2\dots m_{k}}}x^m x^{m_1}\dots x^{m_k}=0.
\eea
The eigenvalues (\ref{SphTensEigen}) are often written as 
\bea\label{SphTensEigenOne}
\la_T=\ell (\ell+s-1)-2,\quad \ell=2,3,\dots
\eea
This completes our discussion of scalar, vector, and tensor spherical harmonics. The eigenvalues are given by (\ref{ScalHarmOneEigen}), (\ref{SphVecEigenOne}), (\ref{SphTensEigenOne}), and the explicit form of the eigenfunctions is determined by solving algebraic equations (\ref{SphHarmScalWave}), (\ref{SphVecConstr}), (\ref{SphTensConstr}).

\appendix

\end{document}